\documentclass[twocolumn,aps,floatfix,superscriptaddress,longbibliography]{revtex4-1}
\usepackage{amsmath,amssymb,eucal,graphicx,float,epstopdf,xparse,xcolor}
\usepackage{epsfig,subfigure}
\usepackage[utf8]{inputenc}
\usepackage{xcolor}

\def\XXint#1#2#3{{\setbox0=\hbox{$#1{#2#3}{\int}$}
     \vcenter{\hbox{$#2#3$}}\kern-.5\wd0}}

\usepackage{xcolor}

\NewDocumentCommand{\eulerian}{omm}
 {
  \genfrac<>{0pt}{}{#2}{#3}
  \IfValueT{#1}{_{\!#1}}
 }

\usepackage[colorlinks=true, urlcolor=blue, anchorcolor=blue, citecolor=blue,filecolor=blue,linkcolor=blue,menucolor=blue]{hyperref}

\begin{document}

\title{Statistics of leaves in growing random trees}

\author{Harrison Hartle}
\email{hhartle@santafe.edu}
\affiliation{Santa Fe Institute, Santa Fe, NM 87501, USA}
\author{P. L. Krapivsky }
\affiliation{Santa Fe Institute, Santa Fe, NM 87501, USA}
\affiliation{Boston University, Boston, MA 02215, USA}

\begin{abstract} 
Leaves, i.e., vertices of degree one, can play a significant role in graph structure, especially in sparsely connected settings in which leaves often constitute the largest fraction of vertices. We consider a leaf-based counterpart of the degree, namely, the leaf degree---the number of leaves a vertex is connected to---and the associated leaf degree distribution, analogous to the degree distribution. We determine the leaf degree distribution of random recursive trees (RRTs) and trees grown via a leaf-based preferential attachment mechanism that we introduce. The RRT leaf degree distribution decays factorially, in contrast with its purely geometric degree distribution. In the one-parameter leaf-based growth model, each new vertex attaches to an existing vertex with rate $\ell+a$, where $\ell$ is the leaf degree of the existing vertex, and $a>0$. The leaf degree distribution has a powerlaw tail when $0<a<1$ and an exponential tail (with algebraic prefactor) for $a>1$. The critical case of $a=1$ has a leaf degree distribution with stretched exponential tail. We compute a variety of additional characteristics in these models and conjecture asymptotic equivalence of degree and leaf degree powerlaw tail exponent in the scale free regime. We highlight several avenues of possible extension for future studies.
\end{abstract}

\maketitle

\section{Introduction}
\label{sec:intro}

Trees are undirected connected graphs without self-loops, multiple edges, or cycles \cite{Drmota}. Mathematical models of random trees have long been subject to research \cite{SCHWENK197771,wilf1981uniform,aldous1990random}; among those, a widely studied class are tree {\it growth} models, characterized by a sequence of stochastic attachment events \cite{Rudas2008}. Random recursive trees (RRTs) are a paradigmatic parameter-free model of growing random trees \cite{Newman,book,Frieze,Hofstad,DM22}: vertices arrive one by one, each attaching to a randomly chosen existing vertex. Another popular mechanism of network growth is preferential attachment (PA) \cite{Simon,barabasi99,barabasi02,DM03}. Many natural network growth mechanisms including PA can be obtained by mild adjustment of the RRT attachment rule \cite{KR01,KR24}. Beyond mathematical appeal, probabilistic models of trees are motivated by numerous real world systems exhibiting treelike patterns of data \cite{ztmk-glc4}.

The {\it degree} of a vertex, that is, its number of neighbors \cite{wilson1979introduction}, is a useful local characteristic of graphs \cite{Diestel} and the simplest measure of centrality \cite{bloch2023centrality}, quantifying the overall level of involvement of a vertex with the rest of the graph; the associated degree distribution is a widely examined quantity in random graph models and data \cite{Newman}, providing a concise description of how the local density of edges varies throughout the graph. In sparse graphs with many {\it leaves} (vertices of degree one), another useful local characteristic, albeit much less frequently examined, is the {\it leaf degree}: the number of neighbors of a vertex that are leaves. Analogously to the degree distribution, the {\it leaf degree distribution} is the fraction of vertices with a given value of leaf degree.

Herein, we focus on properties of growing random trees related to leaves, specifically in RRTs and a leaf-based preferential attachment model wherein leaf degree, rather than degree, drives the formation of edges. We are mostly interested in the limit of large graphs, but we present several exact results valid at all finite sizes. RRTs are well-explored; see books \cite{Newman,book,Frieze,Hofstad,DM22} and \cite{Pittel94,KR01,KR24,KR02-fluct,KR02,Janson05} for a sample of references. The leaf-based growth model we introduce is closely related to widely studied degree-based preferential attachment models; we provide parameter-matched model comparisons to ground this work in established network science theory. Our results reveal a tractable formalism that is applicable to sparse random graphs far beyond the growing tree models considered herein, and provide a foundation for the study of leaf-based statistics and mechanisms in random graphs and data.

\subsection{Notation and conventions}
\begin{figure}[t]
\includegraphics[width=0.7\textwidth,trim=400 200 0 250,clip]{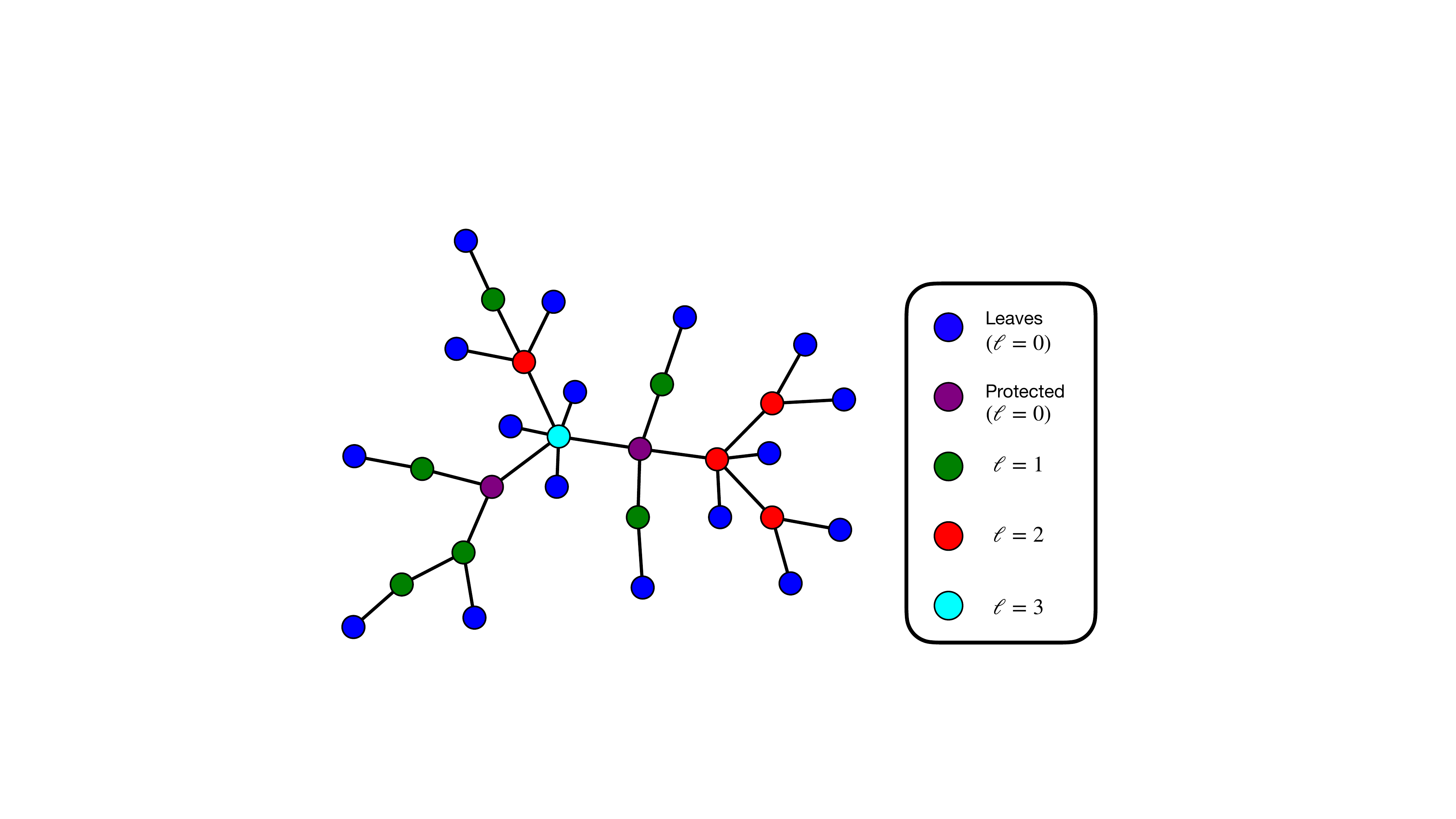}
\caption{A random recursive tree with $N=30$ vertices. The leaf degrees in this tree are $\ell=0,1,2,3$. The number of leaves is $17$ (blue) and the number of protected vertices is $2$ (purple), together constituting $M_0=19$. The numbers of vertices of leaf degree $\ell=1,2,3$ are $M_1=6$ (green), $M_2=4$ (red), and $M_3=1$ (cyan). The ordinary degree values appearing are $k=1,2,3,4,5,6$; the numbers of vertices with these degrees are $(N_1,N_2,N_3,N_4,N_5)=(17,5,4,2,1,1)$. }
\label{fig:Graph_30}
\end{figure}

We consider sequentially grown trees of $N$ vertices labeled $j=1,...,N$ by arrival time (a.k.a., increasing trees \cite{bergeron1992varieties}). The degree $k_i$ of vertex $i$ is its total number of neighbors. $N_k$ denotes the number of vertices of degree $k$ ($k=1,...,N-1$). The total degree is $\sum_{i=1}^Nk_i=2(N-1)$. A {\it leaf} is a vertex of degree one; the number of leaves is $N_1$. The {\it leaf degree} $\ell_i$ of vertex $i$ is the number of leaves among its neighbors, with total leaf degree $\sum_{i=1}^N\ell_i=N_1$. For any vertex $i$, $0\le \ell_i<k_i$, with the exception of the central vertex in a star graph, for which $k_i=\ell_i=N-1$. $M_\ell$ denotes the number of vertices of leaf degree $\ell$ ($\ell=0,...,N-1$); see Fig.~\ref{fig:Graph_30}. In the models we consider, $N_k$ at any $k$ and $M_\ell$ at any $\ell$ each grow proportionally to $N$; we denote the associated intensive quantities as $n_k=N_k/N$ and $m_\ell=M_\ell/N$. Self-averaging of $N_k$ and $M_\ell$, which we argue holds for the models considered herein, implies that $n_k$ and $m_\ell$ converge in probability to their expected values, so we treat the values of $(m_\ell)_{\ell\ge 0}$ and $(n_k)_{k\ge 1}$ as nonrandom. Normalization entails $\sum_{\ell\ge 0} m_\ell=\sum_{k\ge 1}n_k=1$, and the additional sum rules are $\sum_{k\ge 1}kn_k=2$ and $\sum_{\ell\ge 0}\ell m_\ell=n_1$, where $n_1$ is the leaf fraction. Nonleaf vertices are called {\it protected} if they have no leaf neighbors; note that $M_0$ is the total number of leaves plus the total number of protected vertices $P$, so $M_0=N_1+P$. The intensive protected fraction $p=P/N$ satisfies $p=m_0-n_1$. The fraction of vertices that are rank-$1$ (neighbors of leaves) is $\sum_{\ell>0}m_\ell=1-m_0$.

\subsection{Overview of main results}

In Sec.~\ref{sec:degree}, we study the leaf degree for the RRT, assuming that the number $M_\ell$ of vertices with $\ell$ leaves is an extensive and asymptotically self-averaging random quantity. Thus the average ${\mathbb E}(M_\ell) = N m_\ell$ provide the chief insight about the leaf degree distribution in the $N\to\infty$ limit. Herein, we show that for RRTs,
\begin{align}
\label{m-sol}
m_\ell = \int_0^1 dt\,e^{-t}\,\frac{t^\ell}{\ell!}=\frac{\gamma(\ell +1, 1)}{\Gamma(\ell +1)},
\end{align}
with lower incomplete gamma function $\gamma(a,b):=\int_0^ae^{-t}t^{b-1}dt$ and $\Gamma(b)=\gamma(\infty,b)$. The integral representation in \eqref{m-sol} is useful in calculations; $m_\ell$ decay factorially
\begin{equation}
\label{m:asymp}
m_\ell \simeq \frac{e^{-1}}{(\ell+1)!}\qquad\text{when}\quad \ell\gg 1,
\end{equation}
from which we deduce the growth law  
\begin{equation}
\label{ell-max}
\ell_\text{max} \simeq \frac{\log N}{\log(\log N)}
\end{equation}
of the maximal leaf degree (see Appendix~\ref{app:extrema}). The distribution $(m_0,m_1,...)$ represents the probability that a random vertex has leaf degree $\ell$; this is distinct from the leaf degree distribution {\it among nonleaves}, in which case $m_0$ would no longer account for leaves (which themselves have zero leaf neighbors), instead only representing the fraction of vertices that are {\it protected} (nonleaves with leaf degree zero) \cite{hartle2025growing}. Protected vertices have been studied for RRTs and several other classes of random graphs \cite{Prodinger12,Bona14,Devroye14,Mahmoud15,Janson15,Prodinger17, Pittel17,Bona21}. Using $m_0=1-e^{-1}$ which follows from \eqref{m-sol} and the fact that half of the vertices of the large RRT are leaves, the fraction $p=m_0-\frac{1}{2}$ of vertices that are protected is
\begin{equation}
\label{protect:RRT}
p=\frac{1}{2}-\frac{1}{e} = 0.132\,120\,558\,828\ldots
\end{equation}
in the large RRT. Although \eqref{protect:RRT} is well-known \cite{Mahmoud15}, we provide a couple of different derivations. 

\begin{figure}[t]
\includegraphics[width=0.22\textwidth]{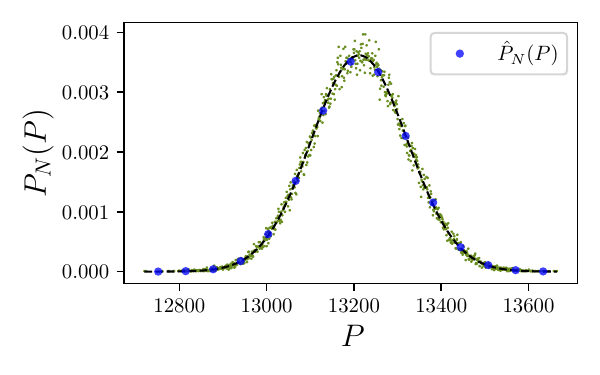}
\includegraphics[width=0.22\textwidth]{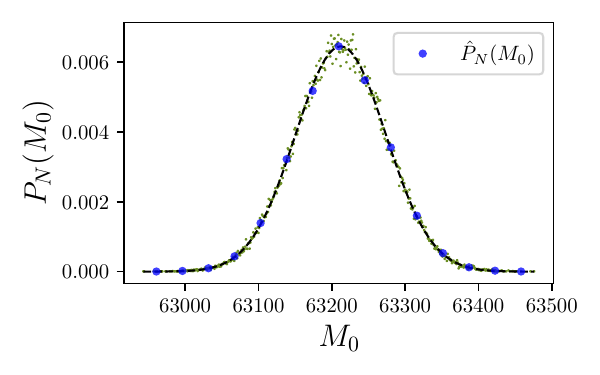}\\
\includegraphics[width=0.22\textwidth,trim=0 20 0 0]{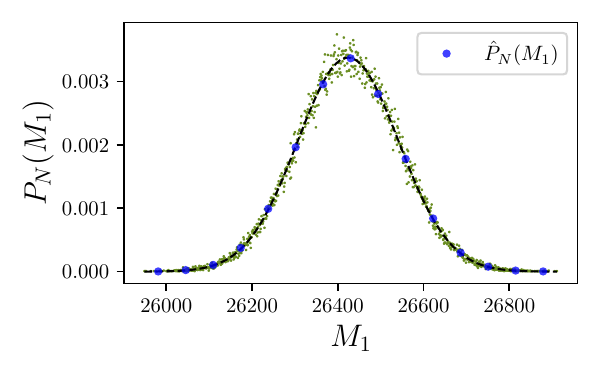}
\includegraphics[width=0.22\textwidth,trim=0 20 0 0]{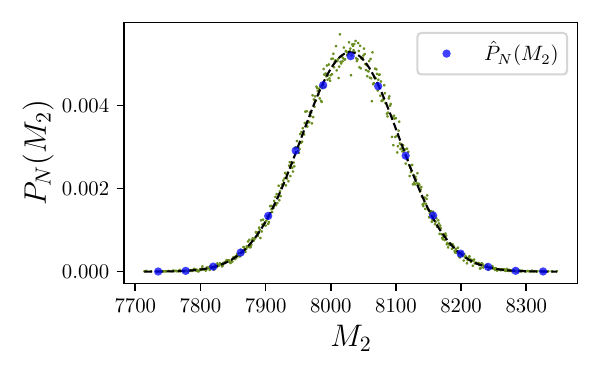}
\caption{Normality of the number of protected vertices $P=M_0-N_1$ and of the numbers of vertices $M_\ell$ with leaf degree $\ell$ for $\ell=0,1,2$. Data from $1.2\times 10^5$ stochastic realizations of the RRT at size $N=10^5$. The intensive variance parameters of Eq.~\eqref{ML-Gauss} are fitted as $\nu_0\approx 0.3816$, $\nu_1\approx0.1396$, and $\nu_2\approx 0.0570$. The green dots are empirical histogram values, blue dots are a binned histogram, and the black dashed lines are fitted normals.}
\label{fig:normality}
\end{figure}

Our calculations in Sec.~\ref{sec:degree} rely on the asymptotic self-averaging of the random quantities $M_\ell$. This is anticipated and supported by simulations that also indicate that the variance ${\mathbb V}(M_\ell)$ is asymptotically extensive:
\begin{equation}
\label{m-nu}
{\mathbb E}(M_\ell) = N m_\ell, \qquad {\mathbb V}(M_\ell) = N \nu_\ell.
\end{equation}
Simulations also suggest that the probability distribution of $M_\ell$ is asymptotically Gaussian:
\begin{equation}
\label{ML-Gauss}
P_N(M_\ell) \simeq \frac{1}{\sqrt{2 \pi N \nu_\ell}}\,\exp\!\left\{-\frac{(M_\ell-N m_\ell)^2}{2 N\nu_\ell}\right\};
\end{equation}
see Fig.~\ref{fig:normality}. Analytically confirming \eqref{m-nu}--\eqref{ML-Gauss} is challenging. Asymptotic normality has been established in many settings including for numerous properties of RRTs \cite{dobrow1996distribution,gopaladesikan2014asymptotic,Janson15}, preferential attachment models \cite{Resnick_Samorodnitsky_2016,wang2017asymptotic,baldassarri2021asymptotic}, and other graphs \cite{10.1214/07-AAP478,BOLLOBAS201253,baa104c6-97f2-3ba5-99bc-2f8bb84a885a}. Certain random quantities in RRTs exhibit fluctuations that can be comprehensively described: in Sec.~\ref{sec:leaves}, we confirm the analogs of \eqref{m-nu}--\eqref{ML-Gauss} for the total number of leaves. Building upon these results, we establish the exact distribution and compute all cumulants. Notably, the cumulants admit a remarkably simple expression involving Bernoulli numbers, while the probability distribution is related to Eulerian numbers. Furthermore, since the total number of leaves equals $\sum_{\ell\geq 1}\ell M_\ell$, the fact that this (weighted) sum of the random quantities $M_\ell$ satisfies the analogs of \eqref{m-nu}--\eqref{ML-Gauss} in turn supports the validity of \eqref{m-nu}--\eqref{ML-Gauss} for individual $\ell$. Analogous normality results for degree-stratified counts $(N_k)_{k\ge 1}$ including covariance characterization have been established in $G(n,p)$ \cite{Janson05}.

In Sec.~\ref{sec:LPA}, we propose growing network models in which newly introduced vertices preferentially attach to vertices with more leaf neighbors. Specifically, we analyze a one-parameter class of random trees where new vertices attach to existing ones at rate $\ell +a$, where $\ell$ is the leaf degree of the existing vertex and $a>0$. The leaf degree distribution exhibits qualitatively different behaviors depending on whether $a$ is smaller than, equal to, or larger than the critical value $a_c=1$. We show that the asymptotic fraction $n_1=N_1/N$  of vertices that are leaves is
\begin{equation}
n_1(a)=\frac{1}{2}-a+\sqrt{a^2+\frac{1}{4}},
\end{equation}
approaching $1$ as $a\downarrow 0$, the RRT value of $\frac{1}{2}$ as $a\rightarrow\infty$, and the value $g-1\approx 0.618$ at $a=1$, where $g=\frac{\sqrt{5}+1}{2}$ is the golden ratio. We also show that the tail of the leaf degree distribution ($\ell \gg 1$) decays as 
\begin{equation}
\label{m-a:tail}
m_\ell \sim 
\begin{cases}
\ell^{-\lambda(a)}                                   & 0<a<1\\
\ell^{-\frac{1}{4}}\,e^{-2\sqrt{g\ell}}         & a = 1 \\
\ell^{-\lambda(a)}\,a^{-\ell}                     & a>1
\end{cases}
\end{equation}
In the regime $0<a<1$, the powerlaw tail exponent $\lambda(a)=\frac{1+n_1(a)}{1-a}$ varies between $\lambda(0)=2$ to $\lambda(1)=\infty$. This regime of exponents is the same as that of shifted linear preferential attachment \cite{KR01}, which has degree distribution tail exponent $\gamma(\delta)=3+\delta$ for $\delta\in(-1,\infty)$. We conjecture that in leaf-$\mathrm{PA}(a)$, the degree distribution is also powerlaw with the same tail exponent as the leaf degree distribution, at least within the regime $\gamma\in(2,3]$. Under this conjecture, leaf-$\mathrm{PA}(a)$ is strictly more flexible than $\mathrm{PA}(\delta)$ in terms of its degree distribution behavior, since it also permits the $a=1$ stretched exponential and the $a>1$ exponential regime. The critical model ($a=1$) has asymptotic fraction of protected vertices given by
\begin{equation}
\label{protect:LPA}
p = g  e^g \,\Gamma(0,g) -g+1 = 0.067\,943\,109\,792\ldots,
\end{equation}
which is just over half that of the RRT (Eq.~\eqref{protect:RRT}). Degree distributions with similar exponential-to-powerlaw crossover behavior have been reported in growth models with concurrent deletion and addition of vertices \cite{bauke2011topological,ghoshal2013uncovering,budnick2025phase}.

\bigskip

The remainder of this article is organized as follows. In Sec.~\ref{sec:degree}, we analyze the leaf degrees of the RRT, obtaining the distribution $(m_\ell)_{\ell\ge 0}$ (Sec.~\ref{subsec:degree}), the distribution of leaf degree of the primordial vertex (Sec.~\ref{ssec:primordialRRT}), and the age-stratified leaf degree distribution (Sec.~\ref{ssec:agestratifiedRRT}). In Sec.~\ref{sec:leaves}, we characterize the distribution of total leaf count $N_1$ in RRTs, including computation of cumulants (Sec.~\ref{subsec:cumulants}) and characterization of the full distribution (Sec.~\ref{subsec:PD}). In Sec.~\ref{sec:LPA} we introduce a leaf-based preferential attachment model, leaf-$\mathrm{PA}(a)$; we study the critical case $a=1$ in Sec.~\ref{ssec:leafPA1}, and $a\ne 0$ in Sec.~\ref{ssec:leafPAnot1}, including analysis of the primordial leaf degree distribution in Sec.~\ref{ssec:primordial}, and a parameter-matched comparison with degree-based preferential attachment in Sec.~\ref{ssec:comparison}. Appendices provide supplementary calculations, including the age-stratified leaf degree distribution in RRTs (Appendix~\ref{app:rrt_solution}) and leaf-$\mathrm{PA}(\delta)$ (Appendix~\ref{app:stratified-leafPA}), estimation of extrema of leaf degree (Appendix~\ref{app:extrema}), exact calculations of cumulants of total leaf count (Appendix~\ref{app:cumulants}), properties of Eulerien numbers arising in characterizing the total leaf count (Appendix~\ref{app:Eulerian}), and analysis of the primordial leaf degree distribution for leaf-$\mathrm{PA}(a)$ in the $a<1$ regime (Sec.~\ref{app:a_le_1}). We provide concluding discussion in Sec.~\ref{sec:discussion}.

\section{Leaf degrees in random recursive trees}
\label{sec:degree}

Before proceeding to leaf degrees, we recall the RRT degree distribution. Vertices of degree $k$ arise by attachment to vertices of degree $k-1$, hence happening with probability proportional to $N_{k-1}$, with the exception of $k=1$ (leaves) which arrive at rate $1$. As such,
\begin{equation}
    \frac{dN_k}{dN}=\frac{N_{k-1}-N_k}{N}+\delta_{k,1}.
\end{equation}
By extensive scaling, $dN_k/dN=n_k$. The asymptotic fractions $n_k$ thus satisfy recurrence
\begin{equation}
\label{nk:rec}
2n_k=n_{k-1} +\delta_{k,1},
\end{equation}
admitting a neat solution
\begin{equation}
\label{nk:RRT}
n_k = 2^{-k}.
\end{equation}

\begin{figure}[t]
\includegraphics[width=1.05\textwidth,trim=1355 560 0 580,clip]{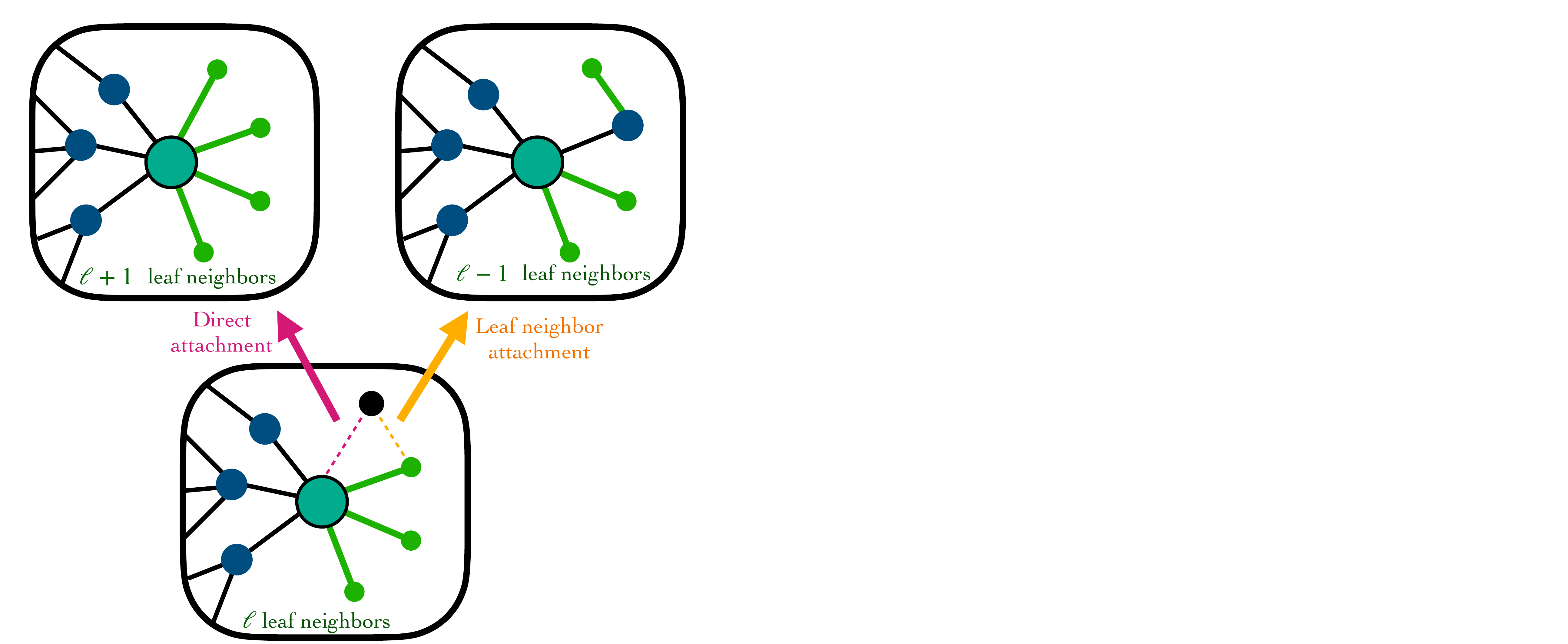}
\caption{Evolution of leaf degree in growing trees. A vertex of leaf degree $\ell$ (turquoise central vertex, lower panel) may either be directly attached to by the arriving vertex (black), or, if $\ell>0$, one of its leaf neighbors (green) may be attached to. Under direct attachment, its leaf degree increases to $\ell+1$ (upper left panel). If one of its leaves is attached to, that previous leaf becomes a nonleaf (blue), yielding a decrease in leaf degree to $\ell-1$ (upper right panel). In the depicted case, $\ell=3$ becomes $\ell=4$ under direct attachment or $\ell=2$ under a leaf neighbor attachment. The total leaf-count goes from $3$ to $4$ in the former case, and is preserved at $3$ in the latter.}
\label{fig:leafdeg-schematic}
\end{figure}

We next turn to the leaf degree distribution, applying similar reasoning to derive rate equations leading to tractable recursions; see Fig.~\ref{fig:leafdeg-schematic}. The leaf degree distribution sheds pale light on large dense networks where the majority of vertices have a leaf degree zero, but for sparse networks, particularly for trees, the leaf degree distribution is useful; it has been considered in several studies, see \cite{leaf01,leaf08} and references therein. In Fig.~\ref{fig:Graph_30}, we depict a tree with $N=30$ vertices, generated by the RRT algorithm. Vertices of different leaf degrees are shown using different colors, with additional distinction between leaves and protected vertices (both having leaf degree zero).

\subsection{Leaf degree distribution of RRT}
\label{subsec:degree}

Let $M_\ell$ denote the number of vertices of leaf degree $\ell$. The normalization condition reads 
\begin{subequations}
\begin{equation}
\label{sum-1}
\sum_{\ell = 0}^{N-1} M_\ell = N.
\end{equation}
Another sum rule is for the total number of leaves, i.e., vertices of degree one,
\begin{equation}
\label{sum-2}
\sum_{\ell=1}^{N-1} \ell M_\ell = N_1.
\end{equation}
\end{subequations} 

We anticipate that the numbers $M_\ell$ are asymptotically self-averaging. The average of $M_\ell$ grows extensively with system size, motivating ansatz $M_\ell=N m_\ell$ for asymptotic fraction $m_\ell$. Counterexamples to this ansatz exist, such as in the scale free leaf-proliferation phenomenon observed in enhanced redirection \cite{GKR13}, isotropic redirection \cite{KR17}, and unlabeled preferential attachment \cite{hartle2025growing}. However, those cases all involve degree and leaf degree distributions with powerlaw tails of exponent $\le 2$, whereas herein all tails decay faster than any powerlaw or as a powerlaw with exponent $>2$. Under the linear scaling ansatz, \eqref{sum-1} reduces to
\begin{subequations}
\begin{equation}
\label{sum-m-1}
\sum_{\ell\geq 0} m_\ell = 1,
\end{equation}
while \eqref{sum-2} becomes
\begin{equation}
\label{sum-m-n1}
\sum_{\ell\geq 1} \ell m_\ell = n_1=\frac{1}{2},
\end{equation}
\end{subequations}
the latter equality following from Eq.~\eqref{nk:RRT}.

\begin{figure}
\includegraphics[width=\linewidth,trim=10 10 0 0,clip]{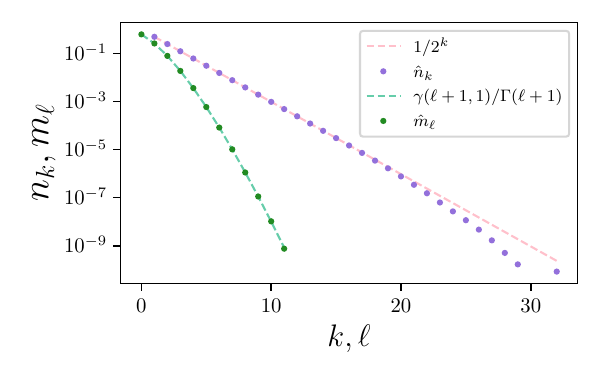}
\caption{Degree and leaf degree distribution in RRTs. Data from $1.2\times 10^5$ stochastic realizations at size $n=10^5$. The degree distribution is $n_k=2^{-k}$ and the leaf degree distribution is given by Eq.~\eqref{m-sol}. Dashed lines represent theoretical curves and dots represent simulation data.}
\label{fig:rrt_dists}
\end{figure}

Under the RRT growth mechanism, the probability of a direct attachment to a vertex of leaf degree $\ell$ is $m_\ell$, and the probability of attachment to a leaf neighbor of a leaf degree $\ell$ vertex is $\ell m_\ell$. $M_\ell$ increases when a vertex of leaf degree $\ell-1$ is attached to, or when a leaf whose unique neighbor has leaf degree $\ell+1$ is attached to; see Fig.~\ref{fig:leafdeg-schematic}. Likewise, $M_{\ell}$ decreases whenever a vertex of leaf degree $\ell$ is attached to, or whenever a leaf whose unique neighbor has leaf degree $\ell$ is attached to Finally, a new leaf, which has leaf degree $0$, arrives at each step. These contributions lead to
\begin{equation}
\label{M-eq}
\frac{d M_\ell}{d N} = \frac{M_{\ell-1} + (\ell+1)M_{\ell+1} - (\ell+1) M_\ell}{N}+\delta_{\ell,0}
\end{equation}
for $\ell\geq 0$. As such, the fractions satisfy the recurrence 
\begin{equation}
\label{m:rec-0}
(\ell+2) m_\ell = m_{\ell-1}+(\ell+1) m_{\ell+1}+\delta_{\ell,0}
\end{equation}
for $\ell\geq 0$. To find the solution of Eqs.~\eqref{m:rec-0}, we introduce the generating function
\begin{equation}
\label{mz:def}
m(z) = \sum_{\ell\geq 0} m_\ell z^\ell,
\end{equation}
and recast the infinite set of equations \eqref{m:rec-0} into a single ordinary differential equation (ODE) for $m(z)$. Multiplying \eqref{m:rec-0} by $z^\ell$ and summing over all $\ell\geq 0$ we arrive at
\begin{equation}
\label{mz:eq}
(1-z)\,\frac{dm}{dz}+(z-2)m+1=0.
\end{equation}
Solving \eqref{mz:eq} subject to $m(1)=1$ implied by the sum rule \eqref{sum-m-1}, we obtain 
\begin{equation}
\label{mz:sol}
m(z) = \frac{1-e^{z-1}}{1-z}.
\end{equation}
Expanding \eqref{mz:sol} in powers of $z$, we arrive at the gamma function expression of Eq.~\eqref{m-sol}. A few explicit expressions for $m_\ell$ are
\begin{equation}
\begin{split}
m_0   &=1 - e^{-1}  \quad    ~        =0.632120558828\ldots \\
m_{1} &= 1-2 e^{-1}      ~~~           = 0.264241117657\ldots\\
m_{2} &= 1-\tfrac{5}{2}\,e^{-1}   ~~= 0.080301397071\ldots\\
m_{3} &= 1-\tfrac{8}{3}\,e^{-1}   ~~= 0.018988156876\ldots \\
m_{4} &= 1-\tfrac{65}{24}\,e^{-1} ~= 0.0036598468273\ldots\\
m_{5} &= 1-\tfrac{163}{60}\,e^{-1} = 0.0005941848176\ldots
\end{split}
\end{equation}
The RRT leaf degree distribution and degree distribution are shown in Fig.~\ref{fig:rrt_dists} alongside simulation data.

\subsection{Primordial leaf degree distribution}
\label{ssec:primordialRRT}

Consider now the primordial (initial) vertex labeled $j=1$, or any one of the first $O(1)$ vertices. We denote by $\pi_\ell$ the probability that the primordial vertex has $\ell$ leaves among its neighbors. For the RRT, $\pi_\ell$ is asymptotically stationary, i.e., independent of $N$, when $N\gg 1$. The probabilities $\pi_\ell$ satisfy the recurrence
\begin{equation}
\label{pi:RRT-rec}
\pi_\ell=\pi_{\ell-1}+(\ell+1)\pi_{\ell+1}-\ell \pi_\ell,
\end{equation}
valid for all $\ell\geq 0$, with convention $\pi_{-1}=0$. Introducing the generating function 
\begin{equation}
\label{pi-z:def}
\pi(z) = \sum_{\ell\geq 0} \pi_\ell z^\ell,
\end{equation}
we recast the recurrence \eqref{pi:RRT-rec} into an ODE
\begin{equation}
\label{pi-z:eq}
\frac{d\pi}{dz} = \pi,
\end{equation}
from which, with $\pi(1)=1$, we obtain $\pi(z) = e^{z-1}$ and hence 
\begin{equation}
\label{pi:RRT}
\pi_\ell=\frac{e^{-1}}{\ell!}.
\end{equation}
Thus, the leaf degree distribution of the primordial vertex is the Poisson distribution of mean $1$, i.e., $e^{-\lambda}\lambda^\ell/\ell!$ at $\lambda=1$. We remark that this is identical to the ordinary degree distribution of the Erd\H{os}--R\'{e}nyi random graph at the percolation point.

The average degree of the primordial vertex, as well as any vertex introduced during the first few steps of the RRT growth process, is on the order of $\log N = k_\text{max} \log 2$. The deviations of the degree of the primordial vertex from the average scale as $\sqrt{\log N}$, see \cite{KR02}. Thus, the probability that the initial vertex is the vertex of the maximal degree approaches zero as $N\to\infty$. The exact distribution of the degree of the initial vertex is also known \cite{KR02}; specifically, it has been expressed in terms of the Stirling numbers \cite{Stirling,Knuth} of the first kind:
\begin{equation}
\label{Stirling}
P(k,N) = \frac{1}{N!}\,{N\brack k}.
\end{equation}

In contrast to the probabilities $P(k,N)$, the probabilities $\pi_\ell$ are stationary, Poissonian, and much simpler than the (non-stationary) probabilities \eqref{Stirling}.

\subsection{Age-stratified RRT leaf degree distribution}
\label{ssec:agestratifiedRRT}

We now explore the dependence of the leaf degree of a vertex on its age. We then compute the average age of the vertex with maximal degree. Mathematically, we want to compute the probabilities $\pi_\ell(j,N)$ that the vertex with label $j$, or age $N-j$, has $\ell$ leaves among its neighbors when the size of the RRT reaches $N$. The probabilities $\pi_\ell(j,N)$ have the following properties:
\begin{itemize}
\item  $\pi_\ell(j,N)\to \pi_\ell$ when $N\to\infty$ and $j$  is kept finite.
\item  $\pi_\ell(j,N)\to \delta_{\ell,0}$ when $N\to\infty$ and $N-j$ is kept finite. 
\item $\pi_\ell(j,N)$ satisfies the sum rule
\begin{equation}
\label{pi:Nj}
\sum_{j=1}^{N} \pi_\ell(j,N) = N m_\ell.
\end{equation}
\end{itemize}
An analogous degree-based age-stratified distribution $c_k(j,N)$ was computed for the RRT and a few other models \cite{KR01}, and it proved useful in elucidating the leadership characteristics of growing networks \cite{KR02}.

For large RRTs ($N\gg 1$), the probabilities $\pi_\ell(j,N)$ exhibit most interesting behavior in the bulk, i.e., when $j\gg 1$ and $N-j\gg 1$. Therefore we can treat $N$ and $j$ as continuous variables and replace differences by derivatives. The probabilities $\pi_\ell(j,N)$ satisfy an infinite set of recurrent partial differential equations (PDEs)
\begin{equation}
\begin{aligned}
\label{pi:RRT-PDE}
N\,\frac{\partial \pi_{\ell}(j,N)}{\partial N} &= \pi_{\ell-1}(j,N)\\
&+(\ell+1)[\pi_{\ell+1}(j,N)-\pi_\ell(j,N)]
\end{aligned}
\end{equation}
for $\ell\geq 0$. We seek the solution $\pi_\ell(j,N)$ as a function of the single variable $x=j/N$ rather than two separate variables $j$ and $N$:
\begin{equation}
\label{pi:ansatz}
 \pi_\ell(j,N)=\Pi_\ell(x), \qquad x = \frac{j}{N}.
\end{equation}

Then \eqref{pi:Nj} becomes $m_\ell = \int_0^1 dx\,\Pi_\ell(x)$, and the PDE, Eq.~\eqref{pi:RRT-PDE}, turns into an ODE
\begin{eqnarray}
\label{pi:RRT-ODE}
-x\,\frac{d \Pi_\ell}{d x} = \Pi_{\ell-1}+(\ell+1)[\Pi_{\ell+1} - \Pi_\ell].
\end{eqnarray}

\begin{figure}
\includegraphics[width=\linewidth,trim=0 12 0 5,clip]{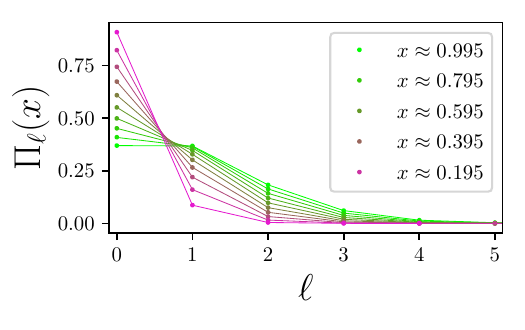}
\caption{Age-stratified distribution of leaf degree at several intermediate normalized ages $x=j/N$. The solid curve shows the theoretical value (Eq.~\ref{Pi:RRT}), and the dots are simulation data from $10^4$ RRTs of size $N=10^4$, with approximation at $x$ taken from indices $Nx-\Delta<j<N x+\Delta$ with $\Delta=50$.} 
\label{fig:age_strat}
\end{figure}

One can check that 
\begin{equation}
\label{Pi:RRT}
\Pi_\ell(x) = \frac{(1-x)^\ell e^{-(1-x)}}{\ell!}
\end{equation}
constitutes the solution of Eqs.~\eqref{pi:RRT-ODE}. See Fig.~\ref{fig:age_strat}. This solution has the properties one would expect, e.g., recovery of Eq.~\eqref{pi:RRT} at $x=0$ ($\Pi_\ell(0)=\pi_\ell$) and $\Pi_\ell(1)=\delta_{\ell,0}$. One could guess \eqref{Pi:RRT} hoping that similarly to \eqref{pi:RRT}, the solution is the Poisson distribution for all $x$; in Appendix~\ref{app:rrt_solution} we provide a derivation. For the leaf-based preferential attachment model considered later (Sec.~\ref{sec:LPA}), we compute the analogous age-stratified quantities in Appendix~\ref{app:stratified-leafPA} for parameter regime $a>1$.

\subsection{Average vertex age as a function of leaf degree}
Using \eqref{Pi:RRT}, we find the average label $J_\ell$ of a vertex with leaf degree $\ell$ is
\begin{equation}
\begin{aligned}
\label{J-ell:def}
\frac{J_\ell}{N} &= \, \frac{\int_0^1 dx\,x \Pi_\ell(x)}{\int_0^1 dx\,\Pi_\ell(x)}
 =1-\frac{\gamma(\ell+2,1)}{\gamma(\ell+1,1)},
 \end{aligned}
\end{equation}
from which
\begin{equation}
\label{J:RRT}
\frac{J_0}{N} = \frac{1}{e-1}\,, \quad \frac{J_1}{N} = \frac{3-e}{e-2}\,, \quad \frac{J_2}{N} = \frac{11-4e}{2e-5}\,,
\end{equation}
etc.; the large $\ell$ behavior of the average label is $J_\ell\simeq N/\ell$, and hence the average label of the vertex with maximal leaf degree is 
\begin{equation}
J_\text{lead} \simeq \frac{N}{\ell_\text{max}} = N\,\frac{\log(\log N)}{\log N},
\end{equation}
derived with Eq.~\eqref{ell-max}; see Appendix~\ref{app:extrema} on the topic of maximal leaf degree.

\section{Leaf count distribution in RRTs}
\label{sec:leaves}

In this section, we examine the distribution of total leaf-count in RRTs. The leaf count is a global quantity, a coarser variable than leaf degree stratified counts $M_\ell$. However, in contrast with the previous and following sections, where we primarily consider average quantities (arguing that they concentrate around their means), here we consider the stochasticity of leaf count. Denote by $\mathcal{L}_N$ the number of leaves in an $N$-vertex RRT. Here we first show that $\mathcal{L}_N$ is a self-averaging extensive random quantity. We then discuss the cumulants of $\mathcal{L}_N$ and its distribution. 

\begin{figure}[t]
\includegraphics[width= 0.47\textwidth,trim=10 20 0 10]{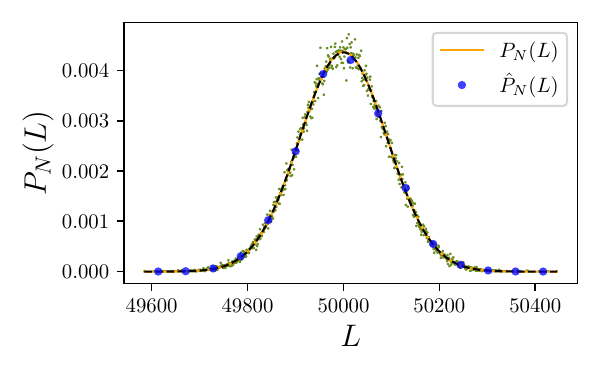}\\
\caption{Normality of total leaf count in RRTs. The orange curve is Eq.~\ref{PLN-Gauss} and the black dashed curve is a fitted normal. Green dots represent empirical probabilities directly, and blue dots represent a binned distribution. Data from $1.2\times 10^5$ graphs of size $N=10^5$.}
\label{fig:normalityL}
\end{figure}

The definition of the RRT leads to the stochastic evolution equation
\begin{equation}
\label{LNN}
\mathcal{L}_{N+1}=
\begin{cases}
\mathcal{L}_N & \text{prob}~~\frac{\mathcal{L}_N}{N}\\
\mathcal{L}_N+1 & \text{prob}~~1-\frac{\mathcal{L}_N}{N}
\end{cases}
\end{equation}
for the total number of leaves $\mathcal{L}_N$ as a random variable. In Appendix~\ref{app:cumulants} we derive the first and second moments,
\begin{equation}
\label{LN-av}
\langle \mathcal{L}_N \rangle = \frac{N}{2}+\frac{1}{N-1},
\end{equation}
and
\begin{equation}
\label{LN2}
\langle \mathcal{L}_N^2 \rangle = \frac{N(3N+1)}{12}+\frac{N}{N-1},
\end{equation}
from which
\begin{equation}
\label{LN-var}
\langle \mathcal{L}_N^2 \rangle_c  = \langle \mathcal{L}_N^2 \rangle - \langle \mathcal{L}_N \rangle^2 = \frac{N}{12}-\frac{1}{(N-1)^2}.
\end{equation}
According to Eq.~\eqref{LN-var}, the mean deviation of $\mathcal{L}_N$ from the average grows as $\sqrt{N}$, i.e., slower than the average. Thus the random variable $\mathcal{L}_N$ is self-averaging. The probability distribution 
\begin{equation}
\label{PLN:def}
P_N(L) = \text{Prob}[\mathcal{L}_N=L]
\end{equation}
is expected to be asymptotically Gaussian \cite{Janson05}, that is, 
\begin{equation}
\label{PLN-Gauss}
P_N(L) \simeq \sqrt{\frac{6}{\pi N}}\exp\!\left\{-\frac{6(L-N/2)^2}{N}\right\},
\end{equation}
reproducing the leading asymptotic behaviors of the average and the variance; see Fig.~\ref{fig:normalityL}. This statement is akin to Eqs.~\ref{ML-Gauss} postulated for $M_\ell$ at each $\ell$.

\subsection{Higher cumulants}
\label{subsec:cumulants}

The average and the variance, \eqref{LN-av} and \eqref{LN-var}, have been computed long ago, see \cite{KR02}. We now show that the expressions for higher cumulants of the number of leaves are also remarkably simple. 

Higher moments $\langle \mathcal{L}_N^p \rangle$ can be recurrently determined similarly to \eqref{LN-av} and \eqref{LN2}. Inspection of the first five cumulants $\langle \mathcal{L}_N^p \rangle_c$ with $p=1,2,3,4,5$ (\eqref{LN-av-app}, \eqref{LN-var-app}, \eqref{LN-3}, \eqref{LN-4}, \eqref{LN-5}) suggests that cumulants with $N\geq p+1$ and arbitrary integer $p\geq 1$ are given by 
\begin{subequations}
\label{conjecture}
\begin{equation}
\label{LN-p}
\langle \mathcal{L}_N^p \rangle_c  = K_p N + \frac{\kappa_p}{(N-1)^p},
\end{equation}
with $K_p$ and $\kappa_p$ determined as follows. In Appendix~\ref{app:cumulants} we obtain the exact values of $\kappa_p$ with $p=1,2,3,4,5$, which suggest the general expression
\begin{equation}
\label{kappa-p}
\kappa_{p} =(-1)^{p-1} (p-1)! 
\end{equation}
for arbitrary $p\geq 1$. The values of $K_p$, which are harder to decipher from values at $p=1,2,3,4,5$, can be expressed through Bernoulli numbers $B_p$:
\begin{equation}
\label{K-B-p}
K_p =\frac{B_p}{p},
\end{equation}
\end{subequations}
which we show in Sec.~\ref{subsec:PD}. Bernoulli numbers $B_p$ are the coefficients in the power series 
\begin{equation}
\label{B:def}
\frac{z}{e^z-1}+z =\sum_{p\geq 0}B_p\,\frac{z^p}{p!}
\end{equation}
Since odd Bernoulli numbers vanish (with one exception, $B_1=\frac{1}{2}$), the conjecture \eqref{LN-p} predicts the vanishing of all higher order odd cumulants. Using \eqref{LN-p} and known values of Bernoulli numbers \cite{Knuth}, we obtain the following leading behavior of the next six even cumulants following \eqref{LN-var}:
\begin{equation}
\begin{aligned}
&\langle \mathcal{L}_N^4 \rangle_c  = -\tfrac{N}{120} \,, \ \ 
\langle \mathcal{L}_N^6\rangle_c = \tfrac{N}{252}\,, \ \ \ \ \ \ \
\langle \mathcal{L}_N^8\rangle_c = - \tfrac{N}{240}\,,\\ 
&\langle \mathcal{L}_N^{10}\rangle_c = \tfrac{N}{132}\,, \ \ \ \
\langle \mathcal{L}_N^{12}\rangle_c = -\tfrac{691 N}{32760}\,, \ \ 
\langle \mathcal{L}_N^{14}\rangle_c = \tfrac{N}{12}.
\end{aligned}
\end{equation}

\subsection{Probability distribution}
\label{subsec:PD}

The evolution rule \eqref{LNN} leads to the recurrence
\begin{equation}
\label{PLN:eq}
P_{N+1}(L+1) = \left(1-\frac{L}{N}\right) P_N(L)+\frac{L+1}{N} P_N(L+1)
\end{equation}
 for the probability distribution \eqref{PLN:def}. The substitution
 \begin{equation}
\label{PQN}
P_{N+1}(L+1)=\frac{2}{N!}Q_N(L),
\end{equation}
recasts Eqs.~\eqref{PLN:eq} into
\begin{equation}
\label{QLN:eq}
Q_N(L) =(N-L) Q_{N-1}(L-1)+ (L+1) Q_{N-1}(L).
\end{equation}
Recalling that $\mathcal{L}_3=2$ we obtain $P_3(2)=1$ and then \eqref{PQN} gives the boundary condition 
\begin{subequations}
\label{BC-IC}
\begin{equation}
\label{IC:Q}
Q_2(1)=1.
\end{equation}
Relations 
\begin{align}
\label{BC:Q}
&Q_N(L)=0  \qquad\text{when}\quad L\geq N\geq 2\\
\label{BC:0}
&Q_N(0)=0  \qquad\text{when}\quad N\geq 2
\end{align}
\end{subequations}
also play the role of the boundary conditions. 

The recurrence \eqref{QLN:eq} is an addition formula for Eulerian numbers \cite{Euler36}, denoted $\eulerian{N}{L}$ \cite{Knuth}. The boundary conditions do not agree with the standard definition of Eulerian numbers, however. Instead, a dual expression can be observed:
\begin{equation}
\label{QQ:Euler}
Q_N(L)+Q_N(N-L-1) = \eulerian{N}{L}.
\end{equation}
In principle, one can determine any $Q_N(L)$ by exploiting the recurrent nature of Eqs.~\eqref{QLN:eq}. See Appendix~\ref{app:Eulerian}.

The numbers $Q_N(L)$ are asymmetric. For $N=6$, for instance, the non-zero numbers $Q_6(L)$ for $L=1,2,3,4,5$ are
\begin{equation*}
16, ~131, ~171, ~41, ~1,
\end{equation*}
The non-zero Eulerian numbers $\eulerian{6}{L}$ for $L=1,2,3,4,5$ are not symmetric,
\begin{equation*}
1, ~57, ~302, ~302, ~57, ~1.
\end{equation*}

In the large $N$ limit, however, both $\eulerian{N}{L}$ and $Q_N(L)$ are sharply peaked around $L=\lfloor N/2\rfloor$, and in that neighborhood the distribution $Q_N(L)$ is asymptotically symmetric: $2Q_N(L)\approx \eulerian{N}{L}$. For the Eulerian random variable $\mathcal{N}_N$
\begin{equation}
\label{nN:def}
\text{Prob}[\mathcal{N}_N=n]= \frac{1}{(N-1)!}\,\eulerian{N-1}{n-1},
\end{equation}
the cumulants are known:
\begin{equation}
\label{N-B}
\langle \mathcal{N}_N^p\rangle_c =\frac{B_p N}{p}
\end{equation}
More precisely, \eqref{N-B} is valid for all integer $p\geq 1$ when  $N>p$, see \cite{David62,Janson13}. This was the source of the guess \eqref{K-B-p} in our case, which we proved by direct pedestrian computations  for $p=1, 2, 3, 4, 5$ (Appendix~\ref{app:cumulants}). The distribution $Q_N(L)$ becomes more and more Eulerian when $N\to\infty$, so the leading behavior of the cumulants predicted by \eqref{LN-p}  is not surprising. What is surprising is that the correction is so small and so simple.
\section{Leaf-based Preferential Attachment Models}
\label{sec:LPA}

In networks growing via the preferential attachment (PA) mechanism \cite{Simon,barabasi99,barabasi02,DM03,DM22,KR01}, a new vertex attaches to an already existing vertex with a probability that is an increasing function of the degree $k$ of the selected vertex. In the simplest PA models, the attachment rate is linear in the degree, for instance, $k+\delta$ with fixed $\delta>-1$. The shift $\delta$ is at least $-1$ because all vertices have degree at least $1$. We denote the latter model as $\mathrm{PA}(\delta)$; it is alternatively known as shifted linear preferential attachment \cite{GR13}, affine preferential attachment \cite{garavaglia2017dynamics}, or preferential attachment with initial attractiveness \cite{dorogovtsev2000structure}. Many variations of preferential have been considered \cite{KR24}, but degree-based rules are the most commonly considered.

Here we analyze a leaf-based preferential attachment mechanism with attachment probability proportional to $\ell+a$, where $\ell$ is the leaf degree of the target vertex and $a>0$. (In contrast with $\delta$ in $\mathrm{PA}(\delta)$, the shift $a$ cannot be negative, since leaf degree can be zero.) The resulting random growing tree, leaf-$\mathrm{PA}(a)$, has qualitatively distinct behavior depending on whether $a>1$, $a=1$, or $a<1$. We begin with the marginal case of leaf-$\mathrm{PA}(1)$, also referred to simply as leaf-$\mathrm{PA}$, which exhibits more subtle behavior than leaf-$\mathrm{PA}(a)$ at $a\ne 1$.

\subsection{Leaf-$\mathrm{PA}(1)$ model}
\label{ssec:leafPA1}

At the critical offset value $a=a_c=1$, the growing tree process is leaf-$\mathrm{PA}(1)$, also referred to simply as leaf-$\mathrm{PA}$, having linear attachment kernel $\ell+1$. Compare to the degree-based model with kernel $k+\delta$ whose critical parameter value $\delta_c=0$ recovers PA, a.k.a., the BA model \cite{barabasi02}. To reconcile $\ell+1$ vs $k$, we note that except for the hub in a star graph, all vertices $i$ in all trees satisfy $k_i\ge \ell_i+1$. As such, if we were to approximate degree by leaf degree, making the approximation $k_i\approx \ell_i+1$ is always an improvement over using $\ell_i$ itself. An offset of exactly $1$ is completely safe of overshooting for any vertex (up to the probability of a star graph), and is correct for all leaves. Hence leaf-$\mathrm{PA}(1)$ is a natural model on its own.

In leaf-PA, the probability of attaching to a vertex of leaf degree $\ell$ is $(\ell+1)M_\ell/K$, where the normalizer $K$ is
\begin{equation}
K=\sum_{\ell\ge 0}(\ell+1)M_{\ell}=N_1+N,
\end{equation}
with $N_1=Nn_1$ denoting the number of leaves. Thus, the analog of Eq.~\eqref{M-eq} reads 
\begin{equation}
\label{M:LPA}
\frac{d M_\ell}{d N} = \frac{\ell M_{\ell-1} + (\ell+1)M_{\ell+1} - (2\ell+1) M_\ell}{N_1+N}+\delta_{\ell,0}
\end{equation}
from which we deduce the recurrence 
\begin{equation}
\label{m:rec-LPA}
(2\ell+2+n_1) m_\ell = \ell m_{\ell-1}+(\ell+1) m_{\ell+1}+(1+n_1)\delta_{\ell,0}
\end{equation}
for $\ell\geq 0$. 
The rate of change of the leaf-count $N_1$ is equal to the probability of attachment to a nonleaf (since attachment to a leaf also removes a leaf). At all $\ell>0$, $M_\ell$ counts the number of nonleaves with leaf degree $\ell$; in contrast, at $\ell=0$, the number of nonleaves is $M_0-N_1$. Thus $N_1$ varies according to
\begin{equation}
\begin{aligned}
\label{N1:LPA}
\frac{dN_1}{dN}&=\frac{\sum_{\ell>0}(\ell+1)M_\ell+M_0-N_1}{K}\\
&=1-\frac{N_1}{N_1+N}
\end{aligned}
\end{equation}

In terms of intensive quantities, \eqref{N1:LPA} becomes 
\begin{equation}
n_1(1+n_1)=1
\end{equation}
from which
\begin{equation}
\label{n1:LPA}
n_1 = \frac{\sqrt{5}-1}{2}=g-1
\end{equation}
where $g=\frac{\sqrt{5}+1}{2}$ is the golden ratio. Eq.~\eqref{m:rec-LPA} becomes
\begin{equation}
\label{m:rec-g}
(2\ell+1+g) m_\ell = \ell m_{\ell-1}+(\ell+1) m_{\ell+1}+g\delta_{\ell,0}.
\end{equation}
To solve Eq.~\eqref{m:rec-g} we again use the generating function $m(z)$ (Eq.~\eqref{mz:def}) and recast the recurrence \eqref{m:rec-g} into a single ODE
\begin{equation}
\label{mz:LPA}
(1-z)^2\,\frac{dm}{dz}-(1-z+g)m+g=0.
\end{equation}
Solving \eqref{mz:LPA} subject to $m(1)=1$ yields
\begin{equation}
\label{mz:sol-LPA}
\begin{aligned}
m(z) = \frac{g}{1-z}\int_0^{1-z}\frac{dw}{w}\,\exp\!\left[\frac{g}{1-z}-\frac{g}{w}\right].
\end{aligned}
\end{equation}

Expanding \eqref{mz:sol-LPA} in powers of $z$ one can determine the fractions $m_\ell$. We used Mathematica to find explicit results for $\ell\leq 10$, from which we arrived at the conjectural general expression 
\begin{equation}
\label{m:Q}
m_\ell = g  e^g \,\Gamma(0,g)\sum_{k=0}^\ell \binom{\ell}{k}\frac{g^k}{k!}-g\, W_\ell(g)
\end{equation}
where 
\begin{equation}
\Gamma(0,g)=\int_g^\infty \frac{du}{u}\,e^{-u}=\text{Ei}(g),
\end{equation}
with $\text{Ei}(g)$ denoting the exponential integral, and with $W_\ell(g)$ being polynomials of $g$ of degree $\ell-1$. Explicit expressions for $W_\ell(g)$ for small $\ell$ are
\begin{equation}
\label{Q-ell}
\begin{split}
&W_0 = 1, ~~ W_1 =1, ~~ W_2 = \tfrac{3}{2}+\tfrac{1}{2}g\\
&W_3 =  \tfrac{11}{6}+\tfrac{4}{3}g+\tfrac{1}{6}g^2\\
&W_4 = \tfrac{25}{12}+\tfrac{29}{12}g+\tfrac{5}{8}g^2+\tfrac{1}{24}g^3\\
&W_5 = \tfrac{137}{60}+\tfrac{37}{10}g+\tfrac{59}{40}g^2+\tfrac{1}{5}g^3+\tfrac{1}{120}g^4\\
&W_6 = \tfrac{49}{20}+\tfrac{103}{20}g+\tfrac{14}{5}g^2+\tfrac{26}{45}g^3+\tfrac{7}{144}g^4 +\tfrac{1}{720}g^5
\end{split}
\end{equation}

\begin{figure}
\includegraphics[scale=0.88,trim = 12 25 0 10,clip]{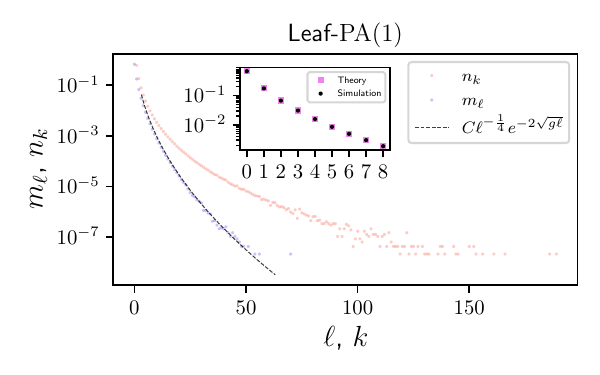}
\caption{ Leaf-$\mathrm{PA}(1)$ leaf degree distribution $(m_\ell)_{\ell\ge 0}$ and degree distribution $(n_k)_{k\ge 1}$ (attachment rate proportional to $\ell+1$). The data are from $480$ graphs at size $N=10^5$. The asymptotic theory curve for $m_\ell$ is Eq.~\eqref{eq:leafPA1_asymp}, with $C\approx 5.3$ a rough numerical estimate. }
\label{fig:dists_leafPA}
\end{figure}

The first three fractions read
\begin{subequations}
\label{m0123}
\begin{align}
\label{m0:LPA}
&m_0 = g\,  e^g \,\Gamma(0,g),\\
\label{m1:LPA}
&m_1 = g(1+g)\,e^g \,\Gamma(0,g) - g,\\
\label{m2:LPA}
&m_2 = g\left(1+2g+\tfrac{g^2}{2}\right) e^g\,\Gamma(0,g)-\tfrac{3g+g^2}{2}.
\end{align}
\end{subequations}
 See the inset of Fig.~\ref{fig:dists_leafPA} which displays theory-simulation agreement for $m_\ell$ at $\ell=0,...,8$. The fraction of protected vertices is $p=m_0-n_1$; using \eqref{m0:LPA} and \eqref{n1:LPA} we arrive at the announced formula \eqref{protect:LPA}. 

The exact expression \eqref{mz:sol-LPA} for the generating function is well-suited for extracting explicit analytical formulae for the fractions $m_\ell$ with small $\ell$, such as Eqs.~\eqref{m0123}. Extracting the tail of the distribution $m_\ell$ from \eqref{mz:sol-LPA} appears challenging; therefore, we outline a WKB approach that allows us to deduce the tail up to a multiplicative constant. We seek the solution of \eqref{m:rec-g} as $e^{-S(\ell)}$, the standard WKB form \cite{Bender}, and obtain
\begin{eqnarray*}
&&\ell\left[e^{S(\ell)-S(\ell+1)}-2+e^{S(\ell)-S(\ell-1)}\right]\\
&&=1+g-e^{S(\ell)-S(\ell+1)}
\end{eqnarray*}
In the $\ell\to\infty$ limit, we use $S(\ell \pm 1)=S \pm S' + \frac{1}{2}S''+\ldots$ where $S=S(\ell), S' = \frac{d S(\ell)}{d\ell}$, etc.  We then expand the exponents and arrive at 
\begin{equation}
\label{S-eq}
(S')^2\ell \simeq g+S'+\ell S''+\ldots
\end{equation}
Thus $S'=\sqrt{g/\ell}$ in the leading order, i.e., $S=2\sqrt{g\ell}$. We use this leading-order expression to compute subleading terms: $S'+\ell S''=(\ell S')'=\frac{1}{2}\sqrt{\frac{g}{\ell}}$ in \eqref{S-eq}. Hence a more accurate solution of \eqref{S-eq} is 
\begin{equation}
S' \simeq  \sqrt{\frac{g}{\ell}+\frac{1}{2\ell}\sqrt{\frac{g}{\ell}}}\simeq\sqrt{\frac{g}{\ell}}+\frac{1}{4\ell}
\end{equation}
which is integrated to give
\begin{equation}
S \simeq 2\sqrt{g\ell} + \frac{1}{4}\, \log \ell + \text{const}
\end{equation}
leading to
\begin{equation}
\label{eq:leafPA1_asymp}
m_\ell \simeq  C\,\ell^{-\frac{1}{4}}\,e^{-2\sqrt{g\ell}}
\end{equation}
for $\ell\gg 1$, as was announced in Eq.~\eqref{m-a:tail}. The amplitude $C$ is not fixed by the WKB approach. We roughly estimate $C\approx 5.3$ from simulation data for visualization in Fig.~\ref{fig:dists_leafPA}. Using the exact formula \eqref{mz:sol-LPA} for the generating function, one can hope to extract the amplitude from the asymptotic behavior in the $z\rightarrow 1$ limit.

\subsection{Leaf-$\mathrm{PA}(a)$ with $a\ne 1$}
\label{ssec:leafPAnot1}

For the leaf-$\mathrm{PA}(a)$ model with arbitrary $a>0$, the probability of attaching to a vertex of leaf degree $\ell$ is $(\ell+a)M_\ell/K(a)$, with normalizer $K(a)$ given by
\begin{equation}
K(a) = \sum_{\ell\ge 0}(\ell+a)M_\ell = N_1+aN.
\end{equation}
The analog of Eq.~\eqref{m:rec-LPA} then reads 
\begin{eqnarray}
\label{m:rec-LPA-a}
[\ell(1+a)+2a+n_1] m_\ell = (\ell-1+a) m_{\ell-1} &&\nonumber \\
+\,a(\ell+1) m_{\ell+1} + (a+n_1)\delta_{\ell,0} \ \ &&
\end{eqnarray}
for $\ell\geq 0$. The fraction of leaves evolves as

\begin{equation}
\begin{aligned}
\frac{dN_1}{dN}&= \frac{\sum_{\ell>0}(\ell+a)M_\ell+a(M_0-N_1)}{K(a)}\\
&= \frac{(1-a)n_1+a}{n_1+a},\\
\end{aligned}
\end{equation}
so, similar to \eqref{N1:LPA} for leaf-$\mathrm{PA}(1)$,
\begin{equation}
\label{n1-a}
n_1(a) = \frac{1}{2}-a+\sqrt{a^2+\frac{1}{4}}.
\end{equation}

\begin{figure}[t]
\includegraphics[width=0.45\textwidth,trim=13 25 0 0]{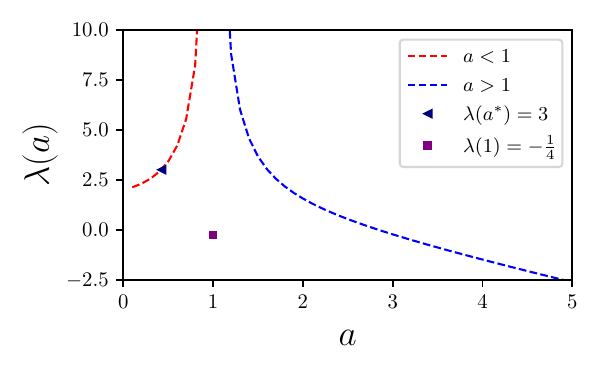}
\caption{The algebraic exponent $\lambda$ versus $a$. The analytical expression for the exponent, Eq.~\eqref{lambda-a}, shows that the exponent diverges as $a$ approaches $1$ from either side.  At $a=a^*$ the exponent equals $3$.  At $a=1$ the algebraic factor is $\ell^{-1/4}$.}
\label{fig:L-a}
\end{figure}

\begin{figure}
    \centering
    \includegraphics[width=1.0\linewidth,trim=0 10 0 0]{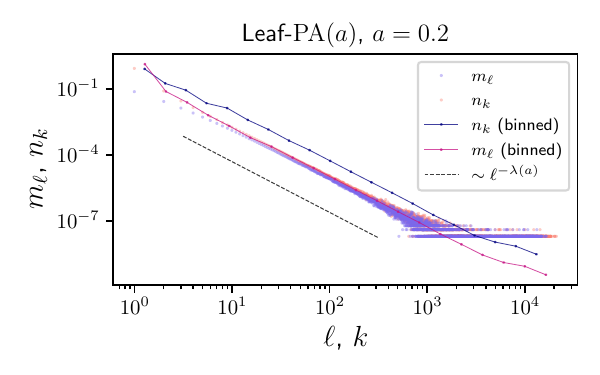}
    \caption{Degree distribution $(n_k)_{k\ge 1}$ and leaf degree distribution $(m_\ell)_{\ell\ge 0}$ in the leaf-$\mathrm{PA}(a)$ model at arbitrarily chosen $a<1$, namely, $a=0.2$. The data are from $480$ graphs at size $N=10^5$; dots represent empirical fractions, dotted lines represent binned data ($20$ bins, logarithmic bin spacing), and the dashed line is powerlaw with theoretical exponent $\lambda(0.2)\approx 2.298...$ from Eq.~\eqref{lambda-a}, having used $n_1(0.2)\approx 0.838...$ by Eq.~\eqref{n1-a}.}    \label{fig:leafPAa0pt2}
\end{figure}

When $a\ne 1$, the WKB treatment of is even simpler than in the critical case of $a=1$ (Sec.~\ref{ssec:leafPA1}). Applying the WKB approach to \eqref{m:rec-LPA-a}, one arrives at the announced tail behavior \eqref{m-a:tail}, namely,
\begin{equation}
\label{m-a:tail}
m_\ell \sim 
\begin{cases}
\ell^{-\lambda(a)}                                   & 0<a<1\\
\ell^{-\frac{1}{4}}\,e^{-2\sqrt{g\ell}}         & a = 1 \\
\ell^{-\lambda(a)}\,a^{-\ell}                     & a>1
\end{cases}
\end{equation}
with algebraic exponent
\begin{equation}
\begin{aligned}
\label{lambda-a}
\lambda(a) &= 
\begin{cases}
\frac{\frac{3}{2}-a+\sqrt{a^2+\frac{1}{4}}}{1-a}           & 0<a<1\\
 2-a + \frac{\frac{3}{2}-a+\sqrt{a^2+\frac{1}{4}}}{a-1} & a>1
\end{cases}\\
&= \begin{cases}
\frac{1+n_1(a)}{1-a}           & 0<a<1\\
 2-a + \frac{1+n_1(a)}{a-1}  & a>1
\end{cases}
\end{aligned}
\end{equation}
with $n_1$ of Eq.~\eqref{n1-a}; see Fig.~\ref{fig:L-a}. Note that the exponent $-\lambda(a)$ in the $a>1$ regime is positive for sufficiently large $a$, but the tail is still dominated by the exponential decay $a^{-\ell}$. The degree and leaf degree distributions of numerically simulated leaf-$\mathrm{PA}(a)$ graphs at $a=0.2$ are shown in Fig.~\ref{fig:leafPAa0pt2}.

It is remarkable that the most natural model, leaf-$\mathrm{PA}(1)$, separates the regimes with an algebraic tail ($a<1$)  and an exponential tail ($a>1$). In the $a\to 1$ limit, $\lambda(a)$ diverges as 
\begin{equation}
\lambda(a) - \frac{g}{|a-1|}=
\begin{cases}
1-\frac{2}{\sqrt{5}} & a<1\\
\frac{2}{\sqrt{5}}     & a>1
\end{cases}
\end{equation}
where we dropped terms of the order of $O(a-1)$. 

To find the exact solution of the recurrence \eqref{m:rec-LPA-a} we recast it into an ODE
\begin{equation}
\label{mz-a}
(1-z)(a-z)\,\frac{dm}{dz}-[a(1-z)+a+n_1]m+a+n_1=0
\end{equation}
for the generating function \eqref{mz:def}. We should solve \eqref{mz-a} subject to the boundary condition $m(1)=1$ following from the normalization requirement, Eq.~\eqref{sum-m-1}.

\begin{figure}[t]
\includegraphics[width=0.6\textwidth,trim=10 100 0 160,clip]{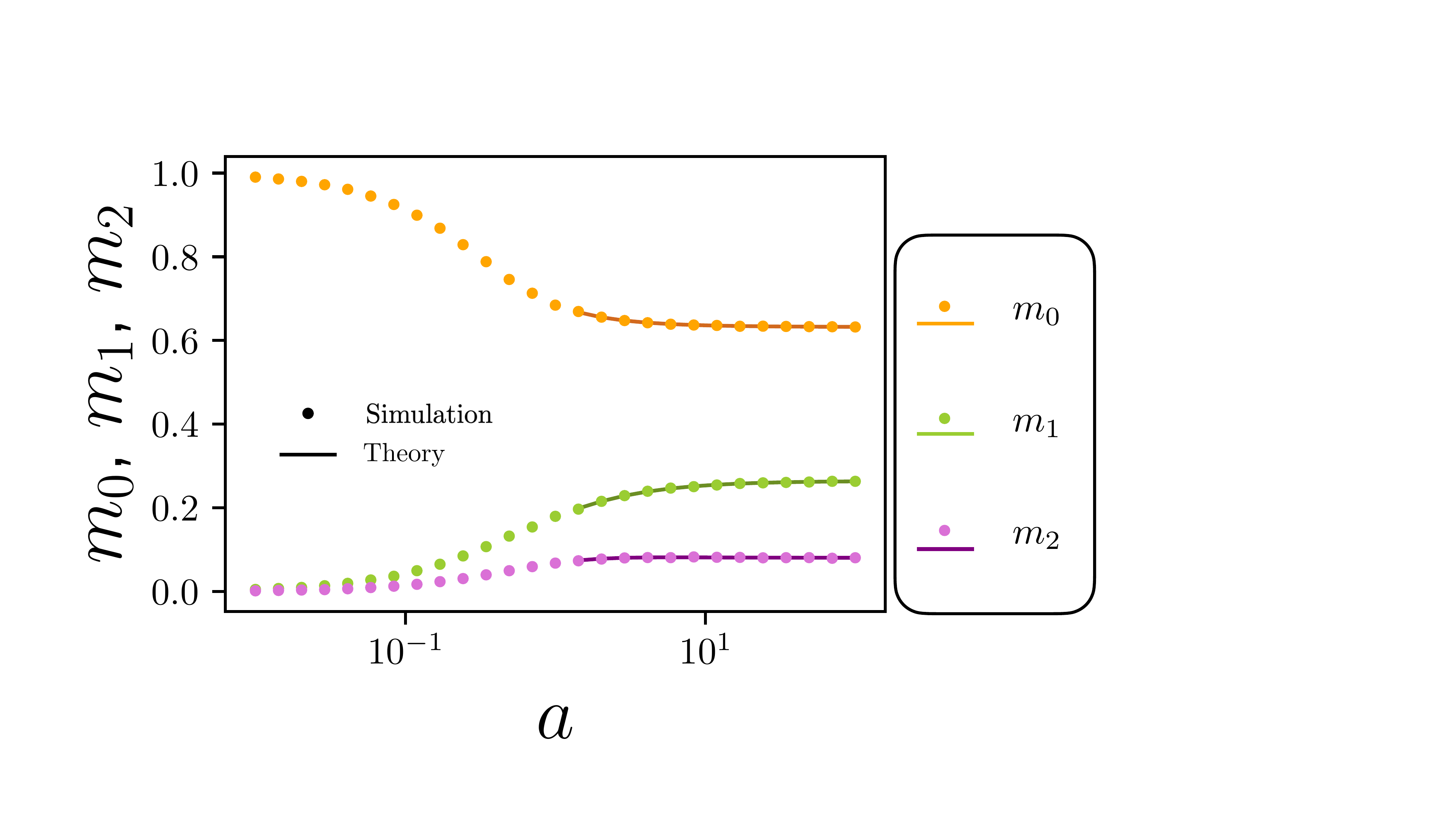}
\caption{The fractions $m_0(a), m_1(a), m_2(a)$ in leaf-$\mathrm{PA}(a)$. Solid curves represent analytical results in the $a>1$ range. The fraction $m_0(a)$ is given by \eqref{m0-a}. An explicit formula for $m_1(a)$ contains two hypergeometric functions; an explicit formula for $m_2(a)$ contains three hypergeometric functions. Dots represent numerical simulation results from $10$ stochastic realizations at size $N=10^4$.}
\label{fig:m012-a}
\end{figure}

When $a>1$, the solution reads

\begin{equation}
\label{mz:sol-LPA-a}
m(z)=(a+n_1)\int_z^1 dw\,\frac{(1-w)^{\mu-1} (a-w)^{\nu-1}}{(1-z)^\mu (a-z)^\nu},\\
\end{equation}
with
\begin{equation}
\label{mu-nu}
\mu = \frac{a+n_1}{a-1}\,, \quad \nu = \frac{a^2-2a-n_1}{a-1}.
\end{equation}
Specializing \eqref{mz:sol-LPA-a} to $z=0$ gives $m(0)=m_0$. One can express the integral via a hypergeometric function:
\begin{equation}
\label{m0-a}
\begin{split}
&m_0(a) = (1-\tfrac{1}{a})^{\frac{a^2-2a-n_1}{a-1}}  \,\Phi(a),\\
&\Phi(a) =F\!\big[\tfrac{n_1+a}{a-1},\tfrac{n_1+3a-a^2-1}{a-1};\tfrac{n_1+2a-1}{a-1};\tfrac{1}{1-a}\big].
\end{split}
\end{equation}
where $n_1=n_1(a)$ is given by \eqref{n1-a}. The formulae for all $m_\ell$ involve hypergeometric functions; the number of hypergeometric functions increases with $\ell$. We do not display these formulae but have plotted $m_1(a)$ and $m_2(a)$, alongside $m_0(a)$ (Eq.~\ref{m0-a}), in Fig.~\ref{fig:m012-a}.

\begin{figure}[t]
\includegraphics[trim=10 20 0 0,width=0.46\textwidth]{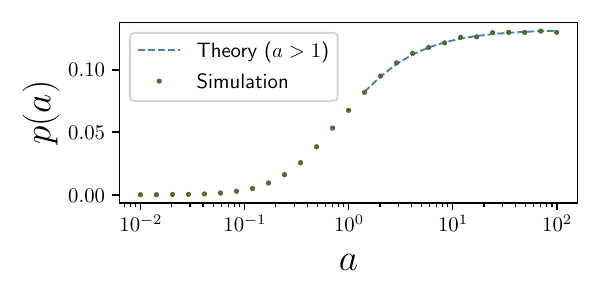}
\caption{The protected fraction, $p(a)=m_0(a)-n_1(a)$, versus $a$. The analytical expressions for $m_0(a)$ (in the range $a>1)$ and $n_1(a)$ are \eqref{m0-a} and \eqref{n1-a}, respectively; the dashed curve represents the analytical curve for $p(a)$.  Dots represent simulation data from $10$ stochastic realizations at size $N=10^4$. For the critical model $a=1$, the value is $p(1) = 0.067\,943...$ (Eq.~\eqref{protect:LPA}). The limiting values are $p(0)=1$ and $p(\infty)=\frac{1}{2}-\frac{1}{e}$.}
\label{fig:protected-a}
\end{figure}

In the $a>1$ range, the fraction of protected vertices, $p=m_0-n_1$, increases from the value \eqref{protect:LPA} at $a=1$ to the value \eqref{protect:RRT} corresponding to $a\rightarrow\infty$ when leaf-$\mathrm{PA}(a)$ approaches the RRT; see Fig.~\ref{fig:protected-a}. Numerical simulations confirm the $a\ge 1$ results and extend the curve to $a<1$.

\subsection{Primordial vertex}
\label{ssec:primordial}

Let $\pi_\ell(N)$ be the probability that the primordial (initial) vertex $j=1$, or any of the first $O(1)$ vertices, has $\ell$ leaves among its neighbors. The probabilities $\pi_\ell(N)$ exhibit different behavior depending on whether $a<1$, $a>1$, or $a=1$. We consider the latter two cases in this section, leaving the case of $a<1$ to Appendix~\ref{app:a_le_1}.

\subsubsection{$a>1$}
\label{subsec:above}

The probabilities $\pi_\ell(N)$ are asymptotically stationary when $a>1$, so we write $\pi_\ell:=\pi_\ell(\infty)$ for short. The same probabilities apply to any of the first $O(1)$ vertices. The probabilities $\pi_\ell$ satisfy the recurrence
\begin{equation}
\label{pi:LPA-rec}
(\ell-1+a)\pi_{\ell-1}+a(\ell+1)\pi_{\ell+1}=[\ell(a+1)+a]\pi_\ell
\end{equation}
valid for all $\ell\geq 0$, upon the convention $\pi_{-1}=0$. Using the generating function $\pi(z)$ of Eq.~\eqref{pi-z:def} we reduce the recurrence \eqref{pi:LPA-rec} to an ODE which we solve subject to the normalization requirement, $\pi(1)=\sum_{\ell\geq 0}\pi_\ell=1$, to find
\begin{equation}
\pi(z) = \left(\frac{a-1}{a-z}\right)^a
\end{equation}
Expanding this generating function yields 
\begin{equation}
\label{pi-a}
\pi_\ell = \frac{(a-1)^a}{a^{\ell+a}}\,\frac{\Gamma(\ell+a)}{\Gamma(a)\Gamma(\ell+1)}.
\end{equation}

The tail of the leaf degree distribution \eqref{pi-a} of the primordial vertex is
\begin{equation}
\label{pi-a:tail}
\pi_\ell \simeq A(a)\,\ell^{a-1}\,a^{-\ell}, \qquad 
A(a) = \frac{(1-a^{-1})^a}{\Gamma(a)}.
\end{equation}
See Fig.~\ref{fig:comparison} wherein the exact solution (Eq.\eqref{pi-a}) and asymptotic (Eq.~\ref{pi-a:tail}) are confirmed. The tail of the leaf degree distribution of the trees generated by leaf-$\mathrm{PA}(a)$ with $a>1$ has the same $a^{-\ell}$ exponential decay, but a different algebraic prefactor (see Eq.~\eqref{lambda-a}).

\begin{figure}
    \centering
\includegraphics[width=1.02\linewidth,trim = 15 15 0 0]{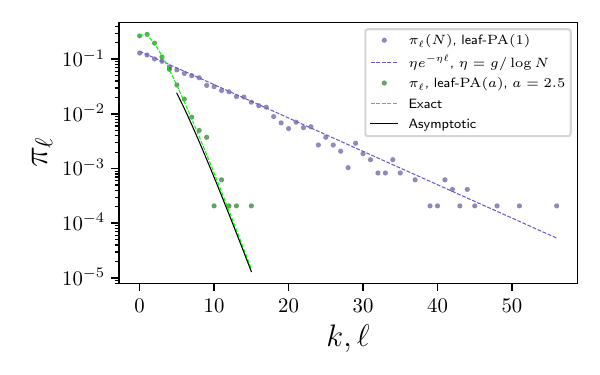}
    \caption{Leaf degree distribution of the primordial vertex in leaf-$\mathrm{PA}$ ($a=1$) and leaf-$\mathrm{PA}(a)$ with $a=2.5$. Simulation data from the first $10$ vertices $j=1,...,10$ in $480$ trees grown to size $N=10^5$. For leaf-$\mathrm{PA(a)}$ at $a>1$, the exact solution (green dashed curve) is Eq.~\eqref{pi-a}, and the asymptotic (black solid curve) is Eq.~\eqref{pi-a:tail}.}
\label{fig:primordial_a1_at1}
\end{figure}

\subsubsection{$a=1$}
\label{subsec:at}

Specializing the recurrence \eqref{pi:LPA-rec} to $a=1$ one finds two independent solutions, $\pi_\ell=1$ and $\pi_\ell=H_\ell$, where $H_n=\sum_{1\leq m\leq n}m^{-1}$ is the Harmonic number. The general solution is $\pi_\ell=C_1+C_2 H_\ell$. This solution is non-normalizable for any choice of the constants $C_1$ and $C_2$. The explanation is simple: stationarity does not hold when $a=1$. The probabilities $\pi_\ell(N)$ satisfy 
\begin{eqnarray}
\label{pi:LPA-rec-1}
\pi_\ell(N+1) &=& \frac{\ell \pi_{\ell-1}(N)+(\ell+1)\pi_{\ell+1}(N)}{gN} \nonumber \\
                    &+&\left[1-\frac{2\ell +1}{gN}\right]\pi_\ell(N)
\end{eqnarray}
We are interested in the large $N$ behavior, and hence we treat $N$ as a continuous variable and replace the difference $\pi_\ell(N+1)-\pi_\ell(N)$ by the derivative. The recurrence \eqref{pi:LPA-rec-1} becomes
\begin{eqnarray}
\label{pi:LPA-1}
gN\,\frac{\partial \pi_\ell(N)}{\partial N} &=& \ell \pi_{\ell-1}(N)+(\ell+1)\pi_{\ell+1}(N)\nonumber \\
                    &-&(2\ell +1)\pi_\ell(N)
\end{eqnarray}

We assume (and confirm a posteriori) that the support of $\pi_\ell(N)$ in $\ell$ diverges with $N$. Hence, we also treat $\ell\gg 1$ as a continuous variable, transform the right-hand side of \eqref{pi:LPA-1}, and arrive at a partial differential equation 
\begin{equation}
\label{pi:LPA-1-PDE}
g\,\frac{\partial \pi_\ell(N)}{\partial \log N} = \frac{\partial }{\partial \ell}\left[\ell \,\frac{\partial \pi_\ell(N)}{\partial \ell}\right].
\end{equation}
The structure of Eq.~\eqref{pi:LPA-1-PDE} suggests that its solution depends on a single scaling variable $\xi=\ell/\log N$. More precisely, we seek a solution of the form
\begin{equation}
\label{pi:F}
\pi_\ell(N) = \frac{1}{\log N}\,F(\xi),
\end{equation}
with prefactor allowing to fulfil the normalization condition that acquires the form 
\begin{equation}
\label{F:norm}
\int_0^\infty d\xi\,F(\xi)=1.
\end{equation}
Substituting the ansatz \eqref{pi:F} into Eq.~\eqref{pi:LPA-1-PDE}, solving an ODE for the scaled function $F(\xi)$, and fixing the amplitude via \eqref{F:norm}, we get
\begin{equation}
\label{pi:LPA-1-sol}
\pi_\ell(N)  
=\eta e^{-\eta\ell}, \ \eta=\frac{g}{\log N}.
\end{equation}
See Fig.~\ref{fig:primordial_a1_at1}.

\subsection{Comparison with $\mathrm{PA}(\delta)$}
\label{ssec:comparison}

Both $\mathrm{PA}(\delta)$ for $-1<\delta<\infty$ and leaf-$\mathrm{PA}(a)$ for $0<a<1$ exhibit powerlaw tails with exponent in range $(2,\infty)$. But this comparison is apples-to-oranges, because the powerlaw is of {\it degree} distribution in $\mathrm{PA}(\delta)$ and of leaf degree distribution in leaf-$\mathrm{PA}(a)$. We conjecture equivalence for sufficiently small powerlaw exponent values; in particular, throughout the classic scale free regime $(2,3]$ for which the 2nd moment diverges. In $\mathrm{PA}$ this regime corresponds to $\delta\in(-1,0]$, whereas in leaf-$\mathrm{PA}(a)$ it corresponds to $a\in(0,a^*]$, with $a^*=1-1/\sqrt{3}$. Under this conjecture, we may compare the models directly with aligned parameters.  The appropriate parameter matching is
\begin{equation} 
\label{eq:a_of_gamma} 
    a =\zeta-\sqrt{\zeta^2-1+\frac{1}{\gamma}},
\end{equation}
where
\begin{equation}
\begin{aligned}\label{eq:zeta_of_gamma}
    \zeta = \frac{\left(\gamma-\frac{3}{2}\right)(\gamma-1)}{\gamma(\gamma-2)}
              =\frac{\left(\delta+\frac{3}{2}\right)(\delta+2)}{(\delta+3)(\delta+1)}, 
\end{aligned}
\end{equation}
with $\gamma=3+\delta$ being the $\mathrm{PA}(\delta)$ degree tail exponent. 

\begin{figure}
\includegraphics[width=0.79\linewidth,trim=10 10 0 0,clip]{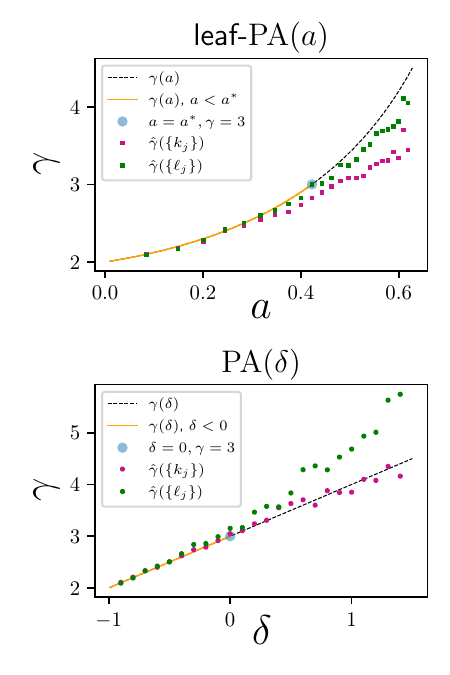}\caption{Estimated tail exponent from degrees and leaf degrees of degree-based and leaf-based preferential attachment, as a function of predicted powerlaw exponent $\gamma$ determining $\delta(\gamma)$ in $\mathrm{PA}(\delta)$ and $a(\gamma)$ in leaf-$\mathrm{PA}(a)$. The estimates all roughly agree in the range $\gamma\in(2,3]$, which we conjecture becomes exact in the thermodynamic limit. Data from $100$ independent realizations of size $N=10^5$ at a range of $a$- and $\delta$-values associated with $\gamma=2.1,...,4.4$.}
\label{fig:comparison}
\end{figure}

Analytically, we neither have the leaf degree distribution $(m_\ell)_{\ell\ge 0}$ for $\mathrm{PA}(\delta)$ nor the ordinary degree distribution $(n_k)_{k\ge 1}$ for leaf-$\mathrm{PA}(a)$, but we conjecture that they are powerlaw-tailed and numerically estimate their tail exponents under parameterization $a(\gamma)$ (Eq.~\eqref{eq:a_of_gamma}) and $\delta(\gamma)=\gamma-3$ for $\gamma>2$. See Fig.~\ref{fig:comparison} which shows the alignment of measured and predicted powerlaw tail exponents: $\lambda(a)$ for $m_\ell$ in leaf-$\mathrm{PA}(a)$, and $3+\delta$ for $n_k$ in $\mathrm{PA}(\delta)$, alongside the estimated exponent for $n_k$ and $m_\ell$ in the two models, respectively; the estimated exponents approximately align in the regime $(2,3]$, supporting our conjecture. The empirically measured exponents compare well to their theoretical counterparts, and degree and leaf degree tail exponent equivalence for $\gamma\le 3$ appears plausible in both models.

\begin{figure}
\includegraphics[width=1\linewidth,trim=0 10 0 -5,clip]{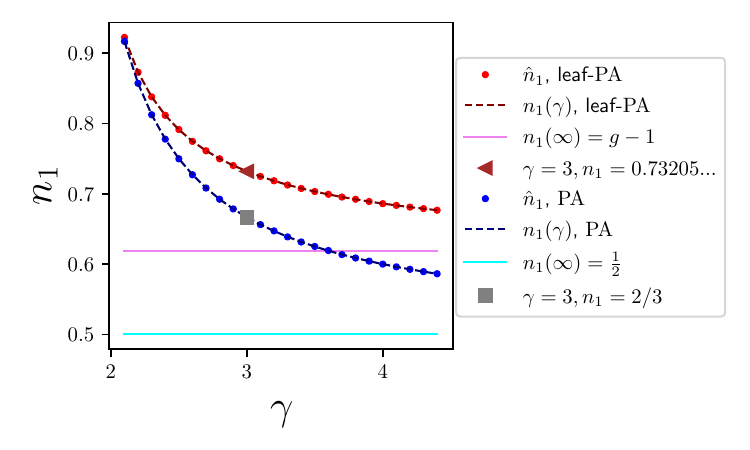}
\caption{$n_1$ in parameter-matched $\mathrm{PA}(\delta)$ and leaf-$\mathrm{PA}(a)$ as a function of tail exponent $\gamma$ with $\delta(\gamma)=\gamma-3$ and $a(\gamma)$ given by Eqs.~\eqref{eq:a_of_gamma},~\eqref{eq:zeta_of_gamma}. Horizontal lines represent the $\gamma\rightarrow\infty$ values of $n_1$, namely, $g-1\approx 0.618...$ and $\frac{1}{2}$, in leaf-PA and PA, respectively. The $\gamma=3$ values are $\frac{2}{3}$ for $\mathrm{PA}(0)$ and $0.732...$ in leaf-$\mathrm{PA}(a^*)$ (Eq.~\eqref{eq:n1_gamma3_leafPA}), respectively.}.
\label{fig:n1_al1}
\end{figure}

A notable special case is that of $\gamma=3$, corresponding to $\delta=0$ in $\mathrm{PA}(\delta)$. To achieve the same value in leaf-$\mathrm{PA}(a)$, we set $a=a^*$, with $a^*$ defined by $\lambda(a^*)=3$, where $\lambda(a)$ is that of Eq.~\eqref{lambda-a} in the regime $0<a<1$. To obtain $a^*$, we take $\gamma=3$ in Eqs.~\eqref{eq:a_of_gamma},~\eqref{eq:zeta_of_gamma} leading to $\zeta=1$ and $a=a^*=1-1/\sqrt{3}$, and asymptotic leaf fraction
\begin{equation}
\label{eq:n1_gamma3_leafPA}
n_1(a^*)=\frac{1}{\sqrt{3}}-\frac{1}{2}+\sqrt{\frac{19}{12}-\frac{2}{\sqrt{3}}}=0.732\,050\ldots,
\end{equation}
in contrast with the leaf fraction $\frac{2}{3}$ of $\mathrm{PA}(0)$; see Fig.~\ref{fig:n1_al1}. The leaf fraction in $\mathrm{PA}(\delta)$ is $n_1(\delta)=(2+\delta)/(3+2\delta)$, with limiting values $n_1(\delta)\rightarrow 1$ as $\delta\rightarrow-1$ and $n_1(\delta)\rightarrow \frac{1}{2}$ as $\delta\rightarrow\infty$.

\begin{figure}
\includegraphics[width=1\linewidth,trim=8 5 0 0,clip]{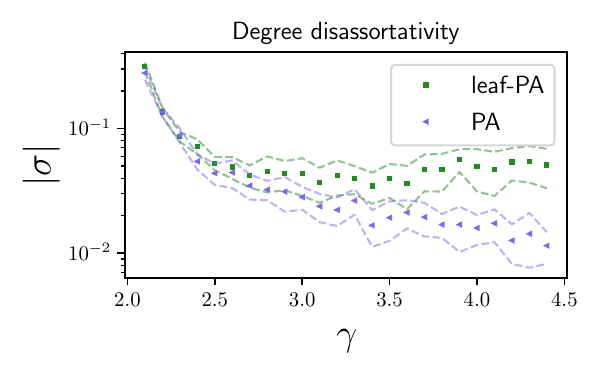}
\includegraphics[width=1\linewidth,trim=0 10 0 0,clip]{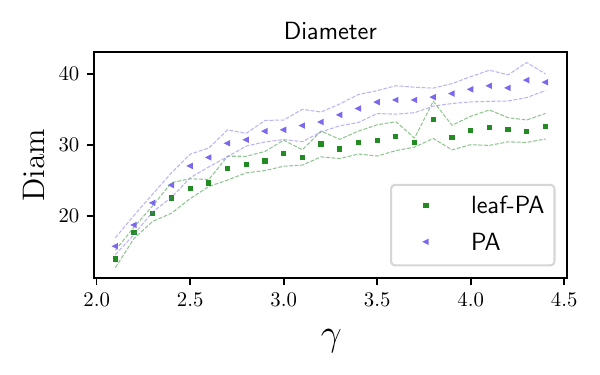}
    \caption{Structural properties in parameter-matched $\mathrm{PA}(\delta)$ and leaf-$\mathrm{PA}(a)$ for $-1<\delta<\infty$ and $0<a<1$, parameterized in terms of the associated powerlaw tail exponent $\gamma$. Across the two models, the properties agree in their general trends, but leaf-PA has consistently higher disassortativity and maximum degree, and consistently lower diameter.}
    \label{fig:comparison2}
\end{figure}

We additionally examine degree correlations and path lengths of leaf-PA and PA in numerical simulations under matched parameters; see Fig.~\ref{fig:comparison2}. We examined the level of {\it degree assortativity} as quantified by the Pearson correlation coefficient, $\sigma=\mathrm{Corr}(k_i,k_j)\in[-1,1]$ for random edge $(i,j)$. We observe that both models are disassortative (negative degree correlation) \cite{Newman}, with the strength of disassortativity $|\sigma|$ strongest at $\gamma=2$, diminishing substantially by $\gamma=3$. Throughout, the strength of disassortativity is larger (i.e., $\sigma$ is more negative) in leaf-PA than PA. The mean {\it diameter} (longest shortest path among all pairs) also exhibits a simple monotonic trend in both models, increasing from minima at $\gamma=2$, and with leaf-PA having a consistently smaller diameter across $\gamma$.

\section{Discussion}
\label{sec:discussion}

In this work, we have investigated the leaf statistics of growing random trees. Analogous to the degree and degree distribution, the leaf degree and leaf degree distribution are useful characteristics of sparse graphs. We have found tractable expressions for leaf statistics of the random recursive tree (RRT), including local structure as captured by the leaf degree distribution and its age-stratified version (Sec.~\ref{sec:degree}), and the distribution of global leaf-count (Sec.~\ref{sec:leaves}). We also introduced a leaf-based preferential attachment model (Sec.~\ref{sec:LPA}), leaf-$\mathrm{PA}(a)$, analogous to degree-based preferential attachment with initial attractiveness, $\mathrm{PA}(\delta)$. The critical model, leaf-$\mathrm{PA}(1)$, has additive shift $a_c=1$, and a stretched exponential tail; for $a>1$ the tail is exponential ($m_\ell\sim a^{-\ell}$ with algebraic prefactor), and for $0<a<1$ the tail is powerlaw ($m_\ell\sim \ell^{-\lambda(a)}$) with exponent $\lambda(a)\in(2,\infty)$. As such, the classic scale free regime $\lambda(a)\in(2,3]$ is accessible via leaf-$\mathrm{PA}(a)$, under the conjecture that within that regime, the degree distribution and leaf degree distribution have the same tail exponents under parameter matching \eqref{eq:a_of_gamma}. Analytical tractability for the exact leaf degree distribution is more challenging in regime $a<1$, yet numerical simulation is straightforward for all values of $a$. As $a\rightarrow\infty$, leaf-$\mathrm{PA}(a)$ approaches the RRT.

These calculations can be widely extended, both for the models considered herein and beyond. This includes previously introduced models, but also many possible new models directly incorporating leaf-based mechanisms such as leaf-$\mathrm{PA}$, for which the leaf-based formalism utilized herein is particularly convenient. Special attention to leaves has been given in certain past studies, such as in graph theory \cite{HENNING2019100,HAYNES2004195,sharada2010domination}---e.g., leaf degree arises in some studies of spanning trees \cite{LIN202597,lin2024existence,ZHOU2024113927}. In comparison with the degree, ubiquitously considered in network science and random graph theory \cite{Newman}, far less consideration has been given to leaf degree \cite{LI2008695}. The leaf degree distribution is thus open for future investigation in sparse random graph models across the board, not just in growing random trees.

The behavior of leaf statistics in real world networks is a notable open subject of interest. Leaves play a key role in real graphs, serving as the outermost remote boundary points of graph structure. For instance, they represent extremely specialist species in food webs \cite{doi:10.1086/285906}, consumers in power grids \cite{deka2016learning}, observed species in phylogenetic trees \cite{rieux2016inferences}, and invariant terminal units of branching tube networks underlying transport processes \cite{west1997}. Leaves also play a key role in various graph algorithms such as greedy leaf removal \cite{zhao2019two}, computation of the $2$-core \cite{bauer2001core,jia2014connecting}, and in the first layers of the onion decomposition \cite{onion}. Furthermore, leaves capture the local symmetry structure of trees, making the leaf-based formalism capable of describing {\it unlabeled} tree growth \cite{hartle2025growing}. We anticipate that the study of leaves and their statistical properties in models and data will provide new insights into sparse network structure and function.

\section*{Acknowledgements}
H.~H.~acknowledges support from the Santa Fe Institute. We thank C.~Moore, Y.~Zhang, and S.~Redner for helpful discussions.

\appendix

\section{Age-stratified leaf degree distribution in RRT}
\label{app:rrt_solution}

From Sec.~\ref{ssec:agestratifiedRRT} we have
\begin{eqnarray}
\label{pi:RRT-ODE-app}
-x\,\frac{d \Pi_\ell}{d x} = \Pi_{\ell-1}+(\ell+1)[\Pi_{\ell+1} - \Pi_\ell],
\end{eqnarray}
and
\begin{equation}
m_\ell=\int_0^1dx\Pi_{\ell}(x).
\label{mellint}
\end{equation}

We also recall definition
\begin{equation}
m(z) = \sum_{\ell\geq 0} m_\ell z^\ell,
\label{mz:def-app}
\end{equation}
and the RRT solution (Eq.~\eqref{mz:sol})
\begin{equation}
m(z) = \frac{1-e^{z-1}}{1-z}.
\label{mz:sol-app}
\end{equation}
Using the generating function 
\begin{equation}
\label{Pi-z:def}
\Pi(x,z) = \sum_{\ell\geq 0} \Pi_\ell(x) z^\ell,
\end{equation}
we reduce an infinite system \eqref{pi:RRT-ODE} to a single PDE
\begin{equation}
\label{Pi-xz:PDE}
-x\,\frac{\partial \Pi}{\partial x}  = (1- z)\left[\frac{\partial \Pi}{\partial z}-\Pi\right].
\end{equation}
Massaging \eqref{Pi-xz:PDE} we transform it into  
\begin{subequations}
\begin{align}
\label{Pi-wave}
&\left(\frac{\partial }{\partial \xi} - \frac{\partial }{\partial \zeta}\right)(e^{-z}\Pi)=0, \\
& \xi=\log x, \quad \zeta = \log(1-z).
\end{align}
\end{subequations}

The general solution of the wave equation \eqref{Pi-wave} reads $\Pi=e^z f(\xi+\zeta)$, or equivalently
\begin{equation}
\label{Pi:sol-gen}
\Pi(x,z) = e^{z} F[x(1-z)].
\end{equation}
With Eqs.~\eqref{mz:def},~\eqref{mellint} we obtain
\begin{equation}
\label{Pi-mz}
m(z)=\int_0^1 dx\,\Pi(x,z).
\end{equation}
To fix the function $F$, we substitute \eqref{mz:sol-app} and \eqref{Pi:sol-gen} into \eqref{Pi-mz}
to yield
\begin{equation*}
\int_0^{1-z}dw\,F(w)=e^{-z}-e^{-1},
\end{equation*}
from which we deduce $F(w)=e^{w-1}$. Thus \eqref{Pi:sol-gen} becomes
\begin{equation}
\label{Pi:sol}
\Pi(x,z) = e^{z+x(1-z)-1}.
\end{equation}
Expanding in powers of $z$ we arrive at \eqref{Pi:RRT} of the main text.

\section{Extremes of leaf degree}
\label{app:extrema}

The maximum degree $k_{\mathrm{max}}$ has been examined in random graph models \cite{RIORDAN_SELBY_2000,BOLLOBAS1980201} including growing random trees \cite{MORI_2005}. From our results on the distribution of leaf degree, we may also extract extrema of leaf degree. In this appendix, we estimate the maximum leaf degree in the $\mathrm{RRT}$, in leaf-$\mathrm{PA}$, and in leaf-$\mathrm{PA}(a)$ for $a\ne 1$.

\subsubsection{$\mathrm{RRT}$}

To estimate the maximal leaf degree $\ell_\text{max}$ of a large tree grown via the RRT procedure, we combine the extreme statistics criterion  
\begin{equation}
\label{crit}
N\sum_{\ell\geq \ell_\text{max}}m_\ell \sim 1,
\end{equation}
with the asymptotic \eqref{m:asymp},
\begin{equation}
m_\ell \simeq \frac{e^{-1}}{(\ell+1)!}\qquad\text{when}\quad \ell\gg 1,
\end{equation}
to obtain growth law  
\begin{equation}
\ell_\text{max} \simeq \frac{\log N}{\log(\log N)}
\end{equation}
of the maximal leaf degree, as announced in Eq.~\eqref{ell-max}.

For comparison, the maximal degree found from a similar criterion $N\sum_{k\geq k_\text{max}}n_k \sim 1$ and the degree distribution \eqref{nk:RRT} exhibits a slightly faster pure logarithmic grows
\begin{equation}
k_\text{max} \simeq \frac{\log N}{\log 2}.
\end{equation}

\subsection{Leaf-$\mathrm{PA}(a)$}

To estimate the maximal leaf degree, we again use the extreme statistics criterion \eqref{crit} together with the asymptotic \eqref{m-a:tail}.

\subsubsection{$a<1$}
\label{app:extrema-leafPA-al1}

If $a<1$, the maximal leaf degree is a non-self-averaging random quantity, with typical magnitude 
\begin{equation}
\label{ell:max}
\ell_\text{max} \sim N^\frac{1}{\lambda(a)-1}
\end{equation}
Both $\ell_\text{typ}$ and $\ell_\text{max}$ exhibit the same scaling with $N$. This can be verified by substituting \eqref{n1-a} into Eq.~\eqref{ell:prim} and \eqref{lambda-a} into Eq.~\eqref{ell:max}.

\subsubsection{$a=1$}
When $a=1$,  the maximal leaf degree is asymptotically self-averaging, and

\begin{equation}
\label{L-max-1}
\ell_\text{max} \simeq  
\frac{(\log N)^2+(\log N)\log(\log N)}{4g}  
\end{equation}

\subsubsection{$a>1$}
Likewise, for $a>1$, $\ell_{\mathrm{max}}$ is self averaging, and
\begin{equation}
\label{L-max}
\ell_\text{max} \simeq  
\frac{\log N}{\log a} - \lambda(a)\,\log(\log N).
\end{equation}

\section{Age-stratified leaf degree distribution $\pi_\ell(j,N)$ in leaf-$\mathrm{PA}(a)$ with $a>1$ }
\label{app:stratified-leafPA}

We briefly look at the probabilities $\pi_\ell(j,N)$ that the vertex with label $j$ has leaf degree $\ell$ in leaf-$\mathrm{PA}(a)$. We consider the $a>1$ range which is more tractable (see Sec.~\ref{subsec:above} and Sec.~\ref{subsec:at}) and the case of $a<1$ in Appendix~\ref{app:a_le_1}. 
The governing equations 
\begin{equation}
\begin{aligned}
\label{pi:LPA-PDE}
(a+&n_1)N\,\frac{\partial \pi_\ell(j,N)}{\partial N} =(\ell-1+a)\pi_{\ell-1}(j, N) \\
&+a(\ell+1)\pi_{\ell+1}(j, N) - [\ell(a+1)+a]\pi_\ell(j,N)
\end{aligned}
\end{equation}
are valid for all $\ell\geq 0$, with convention $\pi_{\ell-1}(j,N)=0$, and derived similarly to Eqs.~\eqref{pi:RRT-PDE} for the RRT. Equations \eqref{pi:LPA-PDE} reduce to Eqs.~\eqref{pi:RRT-PDE} in the $a\to\infty$ limit when the LPA(a) model turns into the RRT. 

Using the ansatz \eqref{pi:ansatz} of  $\pi_\ell(j,N)=\Pi_\ell(x)$, $x =j/N$ we reduce \eqref{pi:LPA-PDE} to
\begin{equation}
\begin{aligned}
\label{pi:LPA-ODE}
-&(a+n_1)x\,\frac{d \Pi_\ell}{d x} =(\ell-1+a)\Pi_{\ell-1} \\&+ a(\ell+1)\Pi_{\ell+1} - [\ell(a+1)+a].\Pi_\ell
\end{aligned}
\end{equation}

The generating function \eqref{Pi-z:def} now satisfies the PDE
\begin{equation}
\label{Pi:LPA-PDE}
-(a+n_1)x\,\frac{\partial \Pi}{\partial x}  = (1- z)\left[(a-z)\frac{\partial \Pi}{\partial z}-a\Pi\right].
\end{equation}

Massaging \eqref{Pi:LPA-PDE} we transform it into a wave equation 
\begin{subequations}
\begin{align}
\label{Pi-wave:LPA}
&\left(\frac{\partial }{\partial \xi} + \frac{\partial }{\partial \zeta}\right)(a-z)^a\Pi = 0,\\
&\xi=\frac{1}{a+n_1}\,\log x, \quad \zeta = \frac{1}{a-1}\,\log\frac{a-z}{1-z},
\end{align}
\end{subequations}
admitting a general solution 
\begin{equation}
\label{Pi-F}
\Pi = (a-z)^{-a} F(y), \quad y = x^\frac{1}{a+n_1} \left(\frac{1-z}{a-z}\right)^\frac{1}{a-1}.
\end{equation}
Substituting \eqref{mz:sol-LPA-a} and \eqref{Pi-F} into \eqref{Pi-mz} we obtain an integral for function $F$:
\begin{eqnarray}
\label{F-eq}
&&\int_0^{ \left(\frac{1-z}{a-z}\right)^\frac{1}{a-1}} dy\,y^{a+n_1-1}F(y) \nonumber \\
&&= \int_z^1 dw\,(1-w)^{\mu-1}(a-w)^{\nu-1}.
\end{eqnarray}
Differentiating \eqref{F-eq} with respect to $z$ we fix $F$ which we combine with \eqref{Pi-F} to yield 
\begin{equation}
\label{Pi-sol:LPA}
\Pi(x,z) = \left[\frac{a-1}{a-z-(1-z)\, x^\frac{1}{\mu}}\right]^a.
\end{equation}
Expanding \eqref{Pi-sol:LPA} in powers of $z$ we arrive at
\begin{equation}
\label{Pi-x:LPA}
\Pi_\ell(x) = \frac{\Gamma(\ell+a)}{\Gamma(a)\Gamma(\ell+1)} \left[\frac{1-x^\frac{1}{\mu}}{a-x^\frac{1}{\mu}}\right]^\ell 
\left[\frac{a-1}{a-x^\frac{1}{\mu}}\right]^a .
\end{equation}

\subsection{Average vertex age as a function of leaf degree }
From Eq.~\eqref{Pi-x:LPA}, we find the average label $J_\ell$ of a vertex with leaf degree $\ell$:
\begin{equation}
\begin{aligned}
\label{J-ell:LPA}
\frac{J_{\ell} }{N}&= \, \frac{\int_0^1 dx\,x \Pi_\ell(x)}{\int_0^1 dx\,\Pi_\ell(x)}\\
&= \frac{\int_0^1 dx\,x\left[\frac{1-x^\frac{1}{\mu}}{a-x^\frac{1}{\mu}}\right]^\ell 
\left[\frac{a-1}{a-x^\frac{1}{\mu}}\right]^a}{\int_0^1 dx\,\left[\frac{1-x^\frac{1}{\mu}}{a-x^\frac{1}{\mu}}\right]^\ell 
\left[\frac{a-1}{a-x^\frac{1}{\mu}}\right]^a},
\end{aligned}
\end{equation}
with $\mu=(a+n_1)/(a-1)$ as in \eqref{mu-nu} of Sec.~\ref{ssec:leafPAnot1}. We deduce the large $\ell$ asymptotic 
\begin{equation}
\label{J-ell:LPA-asymp}
J_{\ell} \simeq N\,\frac{\Gamma(2\mu)}{\Gamma(\mu)}\,\ell^{-\mu}.
\end{equation}
Combining \eqref{J-ell:LPA-asymp} and \eqref{L-max} we find that the average label of the vertex with maximal leaf degree scales as 
\begin{equation}
\bar{j}_\text{lead} \simeq \frac{N}{\ell_\text{max}} \sim \frac{N}{(\log N)^\mu}
\end{equation}
when $a>1$, with $\mu=\frac{1+\sqrt{1+4a^2}}{2(a-1)}$ (see \eqref{n1-a} and \eqref{mu-nu}).

Specializing \eqref{J-ell:LPA} to $\ell=0$ and computing the integrals, we express the average label $J_0$ of leaves and protected vertices through the ratio of hypergeometric functions 
\begin{equation}
\label{J0:LPA}
J_0 = N\,\, \frac{F[a, 2\mu; 2\mu +1; a^{-1}]}{2 F[a, \mu; \mu +1; a^{-1}]}
\end{equation}

There is a single peak: $J_0/N\approx 0.584692$ at $a\approx 2.338629$. When $a\to\infty$, the LPA(a) model reduces to the RRT where  $\frac{J_0}{N}=\frac{1}{e-1}\approx 0.581977$. Specializing \eqref{J-ell:LPA} to $\ell=0$ and taking the $a\downarrow 1$ limit, we express $J_0/N$ through the ratio of exponential integrals:
\begin{equation*}
\begin{aligned}
\frac{J_0}{N}&=\frac{\int_0^1 dx\,x [1-g^{-1}\log x]}{\int_0^1 dx\,[1-g^{-1}\log x]}\\
&=\frac{e^g\,\text{Ei}(-2g)}{\text{Ei}(-g)}=0.580\,965\ldots
\end{aligned}
\end{equation*}
Thus, in the $a>1$ range, $\frac{J_0}{N}$ varies by about $0.6\%$. 

Specializing \eqref{J-ell:LPA} to $\ell=1$ and computing the integrals one finds an exact formula for $J_1/N$ which is more cumbersome than \eqref{J0:LPA}. There is again a single peak:  $J_1/N\approx 0.394542$ at $a\approx 1.1381798$. Specializing \eqref{J-ell:LPA} to $\ell=1$ and taking the $a\downarrow 1$ limit, we obtain
\begin{equation*}
\frac{J_1}{N}=\frac{1+(1+2g)e^{2g}\,\text{Ei}(-2g)}{1+(1+g)e^g \text{Ei}(-g)}=0.394441\ldots
\end{equation*}
Finally, $\frac{J_1}{N}=\frac{3-e}{e-2}\approx 0.392211$ when $a\to\infty$, in agreement with \eqref{J:RRT}.

\section{Analysis of RRT leaf cumulants}
\label{app:cumulants}
We recall the stochastic update equation for RRT leaf count,
\begin{equation}
\label{LNN-app}
\mathcal{L}_{N+1}=
\begin{cases}
\mathcal{L}_N & \text{prob}~~\frac{\mathcal{L}_N}{N}\\
\mathcal{L}_N+1 & \text{prob}~~1-\frac{\mathcal{L}_N}{N}
\end{cases}
\end{equation}
Note that $\mathcal{L}_N$ is deterministic for $N=2$ and $N=3$
\begin{subequations}
\begin{equation}
\label{L23}
\mathcal{L}_2 = \mathcal{L}_3 = 2,
\end{equation}
and genuinely random for $N\geq 4$, e.g.,
\begin{align}
\label{L4}
& \mathcal{L}_4=
\begin{cases}
2 & \text{prob}~~\frac{2}{3} \\
3 & \text{prob}~~\frac{1}{3}
\end{cases}
\\
\label{L5}
& \mathcal{L}_5=
\begin{cases}
2 & \text{prob}~~\frac{1}{3} \\
3 & \text{prob}~~\frac{7}{12} \\
4 & \text{prob}~~\frac{1}{12}
\end{cases}
\end{align}
\end{subequations}

Averaging \eqref{LNN-app} we find
\begin{equation*}
\langle \mathcal{L}_{N+1} \rangle=\langle N^{-1}\mathcal{L}_N^2 \rangle + \langle (\mathcal{L}_N+1)(1-N^{-1}\mathcal{L}_N)\rangle,
\end{equation*}
which simplifies to 
\begin{equation}
\langle \mathcal{L}_{N+1} \rangle=(1-N^{-1})\langle \mathcal{L}_N \rangle + 1.
\end{equation}
Solving this recurrence we obtain
\begin{equation}
\label{LN-av-app}
\langle \mathcal{L}_N \rangle = \frac{N}{2}+\frac{1}{N-1}
\end{equation}
for $N\geq 2$. Similarly from \eqref{LNN-app} we derive
\begin{equation*}
\langle \mathcal{L}_{N+1}^2 \rangle=\langle N^{-1}\mathcal{L}_N^3 \rangle + \langle (\mathcal{L}_N+1)^2(1-N^{-1}\mathcal{L}_N)\rangle,
\end{equation*}
which simplifies to
\begin{equation}
\label{L2-rec}
\langle \mathcal{L}_{N+1}^2 \rangle=\left(1-\frac{2}{N}\right)\langle \mathcal{L}_N^2 \rangle+ \left(2-\frac{1}{N}\right)\langle \mathcal{L}_N \rangle + 1.
\end{equation}
Solving this recurrence we obtain
\begin{equation}
\label{LN2-app}
\langle \mathcal{L}_N^2 \rangle = \frac{N(3N+1)}{12}+\frac{N}{N-1}
\end{equation}
for $N\geq 3$. Using \eqref{LN-av-app} and \eqref{LN2-app} we compute the variance (i.e., the second cumulant):
\begin{equation}
\label{LN-var-app}
\langle \mathcal{L}_N^2 \rangle_c  = \langle \mathcal{L}_N^2 \rangle - \langle \mathcal{L}_N \rangle^2 = \frac{N}{12}-\frac{1}{(N-1)^2}.
\end{equation}
These variances are positive, $\langle \mathcal{L}_N^2 \rangle_c>0$, for $N>3$ reflecting that $\mathcal{L}_N$ are random. The quantity $\mathcal{L}_3$ is deterministic, and \eqref{LN-var-app} indeed gives $\langle \mathcal{L}_3^2 \rangle_c  = 0$. One cannot use the recurrence \eqref{L2-rec} to determine $\langle \mathcal{L}_2^2 \rangle$ since $\mathcal{L}_N$ is defined only for $N\geq 2$. Hence Eqs.~\eqref{LN-var-app} apply to $N\geq 3$. Since $\mathcal{L}_2$ is deterministic, $\langle \mathcal{L}_2^2 \rangle_c  = 0$. Higher moments $\langle \mathcal{L}_N^p \rangle$ can be recurrently determined similarly to \eqref{LN-av-app} and \eqref{LN2-app}. 

\subsubsection{Third cumulant}

The third moment satisfies
\begin{equation}
\begin{aligned}
\langle \mathcal{L}_{N+1}^3 \rangle &= \left(1-\tfrac{3}{N}\right)\langle \mathcal{L}_N^3\rangle\\
&+ 
3\left(1-\tfrac{1}{N}\right)\langle \mathcal{L}_N^2 \rangle + \left(3-\tfrac{1}{N}\right)\langle \mathcal{L}_N \rangle + 1.
\end{aligned}
\end{equation}
Simplifying this recurrence with the help of \eqref{LN-av-app} and \eqref{LN2-app} we obtain
\begin{equation}
\begin{aligned}
\langle \mathcal{L}_{N+1}^3 \rangle &= \left(1-\tfrac{3}{N}\right)\langle \mathcal{L}_N^3\rangle \\
&+ \frac{3N^2+4N+13}{4}+\frac{2}{N-1}+\frac{1}{N},
\end{aligned}
\end{equation}

which is solved to yield
\begin{equation}
\label{LN3}
\langle \mathcal{L}_N^3 \rangle = \frac{N(N^3+5N+2)}{8(N-1)}
\end{equation}
for $N\geq 4$. Using \eqref{LN-av-app}, \eqref{LN2-app}, \eqref{LN3} we compute the third cumulant:
\begin{equation}
\label{LN-3}
\langle \mathcal{L}_N^3 \rangle_c  = \langle \mathcal{L}_N^3 \rangle - 3\langle \mathcal{L}_N^2 \rangle \langle \mathcal{L}_N \rangle
+ 2 \langle \mathcal{L}_N \rangle^3 = \frac{2}{(N-1)^3}
\end{equation}
for $N\geq 4$. Since the quantities $\mathcal{L}_2$ and $\mathcal{L}_3$ are deterministic, $\langle \mathcal{L}_2^3 \rangle_c  = \langle \mathcal{L}_3^3 \rangle_c  = 0$. The third cumulants \eqref{LN-3} are positive for $N\geq 4$, but they quickly approach to zero in the $N\to\infty$ limit.

\subsubsection{Fourth cumulant}
The fourth moment satisfies

\begin{equation}
\begin{aligned}
\langle \mathcal{L}_{N+1}^4 \rangle &=\left(1-\tfrac{4}{N}\right)\langle \mathcal{L}_N^4\rangle + \left(4-\tfrac{6}{N}\right)\langle \mathcal{L}_N^3\rangle \\
&+ \left(6-\tfrac{4}{N}\right)\langle \mathcal{L}_N^2 \rangle  + \left(4-\tfrac{1}{N}\right)\langle \mathcal{L}_N \rangle + 1,
\end{aligned}
\end{equation}
from which
\begin{equation}
\begin{aligned}
\langle \mathcal{L}_{N+1}^4 \rangle& = \left(1-\tfrac{4}{N}\right)\langle \mathcal{L}_N^4\rangle  \\
&+\tfrac{6N^3+15N^2+45N+68}{12} + \tfrac{3}{N-1}+\tfrac{1}{N},
\end{aligned}
\end{equation}
leading to
\begin{equation}
\label{LN4}
\langle \mathcal{L}_N^4 \rangle = \frac{N(15N^4+15N^3+95N^2+113N+2)}{240(N-1)}
\end{equation}
for $N\geq 5$.  For small $N$, we use \eqref{L23}--\eqref{L4} and directly compute $\langle \mathcal{L}_2^4 \rangle  = \langle \mathcal{L}_3^4 \rangle  = 16$ and $\langle \mathcal{L}_4^4 \rangle  =  \frac{113}{3}$. The fourth cumulant reads
\begin{equation}
\label{LN-4}
\langle \mathcal{L}_N^4 \rangle_c  = -\frac{N}{120} -\frac{6}{(N-1)^4}
\end{equation}
for $N\geq 5$. We have again $\langle \mathcal{L}_2^4 \rangle_c  = \langle \mathcal{L}_3^4 \rangle_c  = 0$, and the remaining quantity $\langle \mathcal{L}_4^4 \rangle_c=-\frac{2}{27}$ is extracted from \eqref{L4}. 

\subsubsection{Fifth cumulant}

The fifth moment satisfies

\begin{equation}\begin{aligned}
\langle \mathcal{L}_{N+1}^5 \rangle& = \left(1-\tfrac{5}{N}\right)\langle \mathcal{L}_N^5\rangle\\ 
&+ 5\left(1-\tfrac{2}{N}\right)\langle \mathcal{L}_N^4\rangle + 10\left(1-\tfrac{1}{N}\right)\langle \mathcal{L}_N^3\rangle\\
 &+ \left(10-\tfrac{5}{N}\right)\langle \mathcal{L}_N^2 \rangle  + \left(5-\tfrac{1}{N}\right)\langle \mathcal{L}_N \rangle + 1,
\end{aligned}\end{equation}
from which
\begin{equation}\begin{aligned}
\langle \mathcal{L}_{N+1}^5 \rangle &=\left(1-\tfrac{5}{N}\right)\langle \mathcal{L}_N^5\rangle + \tfrac{4}{N-1}+\tfrac{1}{N} \\
&+\tfrac{15N^4+60N^3+185N^2+388N+368}{48} .
\end{aligned}\end{equation}

Solving this recurrence yields
\begin{equation}
\label{LN5}
\langle \mathcal{L}_N^5 \rangle = \tfrac{N(3N^5+7N^4+25N^3+53N^2+12N-4)}{96(N-1)}
\end{equation}
for $N\geq 6$. The fifth cumulant reads
\begin{equation}
\label{LN-5}
\langle \mathcal{L}_N^5 \rangle_c  = \frac{24}{(N-1)^5}
\end{equation}
for $N\geq 6$. We have again $\langle \mathcal{L}_2^5 \rangle_c  = \langle \mathcal{L}_3^5 \rangle_c  = 0$, while $\langle \mathcal{L}_4^5 \rangle_c$ and $\langle \mathcal{L}_5^5 \rangle_c$ can be extracted from previous results as before. Similarly to the third cumulant, the fifth cumulant \eqref{LN-5} vanishes in the $N\to\infty$ limit.

\section{Eulerian number expressions}
\label{app:Eulerian}
Eulerian numbers \cite{Knuth} are denoted $\eulerian{N}{L}$. One can determine any $Q_N(L)$ by exploiting the recurrent nature of Eqs.~\eqref{QLN:eq}. The expressions for $Q_N(L)$ with fixed small $L$ and arbitrary $N$ are rather simple:
\begin{subequations}
\begin{align}
\label{Q1N}
& Q_N(1)  = 2^{N-2} ,\\
\label{Q2N}
& Q_N(2) = 3^{N-1} - (N+1)2^{N-2}, \\
\label{Q3N}
& Q_N(3) = \tfrac{3}{8}\,4^{N} + (N+1)\big[N\cdot 2^{N-3} - 3^{N-1}\big],
\end{align}
\end{subequations}
etc. Combining such results with \eqref{QQ:Euler} yields $Q_N(L)$ for fixed small $N-L$ and arbitrary $N$:
\begin{subequations}
\begin{align}
\label{QNN}
& Q_N(N-1) =1, \\
\label{QN2N}
& Q_N(N-2) =3\cdot 2^{N-2} - N - 1, \\
\label{QN3N}
&Q_N(N-3) =2\cdot 3^{N-1} -3(N+1) 2^{N-2} \nonumber \\
&\qquad\qquad\qquad  + \tfrac{1}{2}N(N+1), \\
\label{QN4N}
& Q_N(N-4) = \tfrac{3}{8}\,4^{N} - 2(N+1)3^{N-1}  \nonumber \\
&\qquad\qquad\qquad  +  N(N+1)\big[3\cdot 2^{N-3} -  \tfrac{N-1}{6} \big],
\end{align}
\end{subequations}
etc. Specializing \eqref{QLN:eq} to $L=1$ and using \eqref{BC:0} we deduce $Q_N(1)=2Q_{N-1}(1)$ which is iterated to give \eqref{Q1N}. Specializing \eqref{QLN:eq} to $L=2$ and using \eqref{Q1N} we arrive at the recurrence
\begin{equation*}
Q_N(2) = 3 Q_{N-1}(2) + (N-2) 2^{N-3},
\end{equation*}
which is solved to give \eqref{Q2N}. Specializing \eqref{QLN:eq} to $L=3$ and using \eqref{Q2N} we arrive at the recurrence
\begin{equation*}
Q_N(3) = 4 Q_{N-1}(3) + (N-3)[3^{N-2}-N\cdot 2^{N-3}],
\end{equation*}
which is solved to give \eqref{Q3N}. Specializing \eqref{QLN:eq} to $L=N-1$ and using \eqref{BC:Q} we find $Q_N(N-1)=Q_{N-1}(N-2)$ which is iterated to give \eqref{QNN}. Specializing \eqref{QQ:Euler} to $L=1$, using \eqref{Q1N} and the explicit expression \cite{Knuth}
\begin{equation}
\label{Euler-1}
\eulerian{N}{1} = 2^N-N-1,
\end{equation}
we deduce \eqref{QN2N}. Specializing \eqref{QQ:Euler} to $L=2$, using \eqref{Q2N} and the explicit expression \cite{Knuth}
\begin{equation}
\label{Euler-2}
\eulerian{N}{2} = 3^N-(N+1)2^N+\binom{N+1}{2},
\end{equation}
we deduce \eqref{QN3N}. Specializing \eqref{QQ:Euler} to $L=3$, using \eqref{Q3N} and explicit expression \cite{Knuth}
\begin{equation}
\label{Euler-3}
\eulerian{N}{3} = 4^N - (N+1)3^N+2^N\binom{N+1}{2}- \binom{N+1}{3},
\end{equation}
we deduce \eqref{QN4N}. Similarly one can use \eqref{QLN:eq} to recurrently compute $Q_N(5), ~Q_N(6)$, etc. Utilizing \eqref{QQ:Euler} and explicit expressions \cite{Knuth,Kyle} for $\eulerian{N}{4}, ~\eulerian{N}{5}$, etc. one then derives explicit expressions  for $Q_N(N-5), ~Q_N(N-6)$, etc.

\section{Primordial leaf degree distribution in leaf-$\mathrm{PA}(a)$ for $0<a<1$}
\label{app:a_le_1}

For leaf-$\mathrm{PA}(a)$ with $0<a<1$, we begin with the exact recurrence 

\begin{equation}
\begin{aligned}
\label{pi:LPA-rec-a}
\pi_\ell(N+1) &= \left[1-\frac{\ell(1+a) +a}{(a+n_1)N}\right] \pi_\ell(N)\\
&+\frac{(\ell-1+a) \pi_{\ell-1}(N)+a(\ell+1)\pi_{\ell+1}(N)}{(a+n_1)N},
\end{aligned}
\end{equation}
and transform it into
\begin{equation}
\begin{aligned}
\label{pi:LPA-a}
(a+n_1)N\,&\frac{\partial \pi_\ell(N)}{\partial N} = (\ell-1+a)  \pi_{\ell-1}(N) \\
                    &+ a(\ell+1)\pi_{\ell+1}(N)-[\ell(1+a) +a]\pi_\ell(N),
\end{aligned}
\end{equation}
applicable when $N\gg 1$. The span of $\pi_\ell(N)$ in $\ell$ again diverges with $N$, so we treat $\ell\gg 1$ as a continuous variable, we expand the right-hand side of Eq.~\eqref{pi:LPA-a} and arrive at a hyperbolic PDE
\begin{equation}
\label{pi:LPA-a-PDE}
\left[(a+n_1)\,\frac{\partial }{\partial \log N} + (1-a)\frac{\partial }{\partial \log \ell}\right] \ell \pi_\ell(N) = 0
\end{equation}
in the leading order. The wave equation Eq.~\eqref{pi:LPA-a-PDE} suggests that typical leaf degree $\ell_{\mathrm{typ}}$ of the primordial vertex: 
\begin{equation}
\frac{\log N}{a+n_1}\approx \frac{\log \ell_\text{typ}}{1-a},
\end{equation}
and hence
\begin{equation}
\label{ell:prim}
\ell_\text{typ} \sim N^\frac{1-a}{a+n_1}.
\end{equation}
The quantities $\ell_{\mathrm{typ}}$ and $\ell_{\mathrm{max}}$ have the same scaling (see Appendix~\ref{app:extrema-leafPA-al1}).

\end{document}